\documentclass[a4paper,fleqn,usenatbib]{mnras}
\usepackage{graphicx}
\usepackage{amssymb}
\usepackage{amsmath}
\usepackage{booktabs}
\usepackage{varwidth}
\usepackage[T1]{fontenc}
\usepackage{ae,aecompl}
\usepackage{natbib,times}
\usepackage{lscape}
\usepackage{array}
\usepackage{setspace}
\usepackage{helvet}
\usepackage{float}
\newcommand{\PreserveBackslash}[1]{\let\temp=\\#1\let\\=\temp}
\newcolumntype{C}[1]{>{\PreserveBackslash\centering}p{#1}}
\newcolumntype{R}[1]{>{\PreserveBackslash\raggedleft}p{#1}}
\newcolumntype{L}[1]{>{\PreserveBackslash\raggedright}p{#1}}

\begin{document}

\title[More relaxed intracluster gas than galaxies] {More relaxed intracluster gas than galaxies in clusters in quasi-equilibrium}

\author[Z. S. Yuan et al.]
       {Z. S. Yuan,$^{1,2}$ \thanks{E-mail: zsyuan@nao.cas.cn}
        J. L. Han,$^{1,2,3}$ \thanks{E-mail:hjl@nao.cas.cn}
        H. B\"ohringer,$^{4,5}$
        Z. L. Wen,$^{1,2}$
        and G. Chon$^4$
\\
1. National Astronomical Observatories, Chinese Academy of Sciences, 
20A Datun Road, Chaoyang District, Beijing 100101, China\\
2. CAS Key Laboratory of FAST, NAOC, Chinese Academy of Sciences,
           Beijing 100101, China\\
3. School of Astronomy, University of Chinese Academy of Sciences,
           Beijing 100049, China\\
4. University Observatory, Ludwig-Maximilians-Universit\"at M\"unchen, 
           Scheinerstr. 1, 81679 M\"unchen, Germany\\
5. Max-Planck-Institut f\"ur Extraterrestrische Physik, 
           85748 Garching, Germany}

\date{Accepted XXX. Received YYY; in original form ZZZ}

\label{firstpage}
\pagerange{\pageref{firstpage}--\pageref{lastpage}}
\maketitle

%%%%%%%%%%%%%%%%%%%%%%%%%%%%%%%%%%%%%%%%%%%%%%%%%%%%%%%%%%%%%%%%%%%%%%%%%%%%%

\begin{abstract}
During cluster mergers, the intracluster gas and member galaxies undergo dynamic evolution, but at different timescales and reach different states. We collect 24 galaxy clusters in quasi-equilibrium state as
indicated by the X-ray image, and calculate the cluster orientations and
three kinds of dynamical parameters, i.e., the normalized centroid
offset, the sphere index and the ellipticity, for these clusters from
the distributions of member galaxies and also the intracluster gas. We
find consistent alignments for the orientations estimated from the two
components. However, the three kinds of dynamical parameters indicated
by member galaxies are systematically larger than those derived from
the gas component, suggesting that the gas component is more
relaxed than member galaxies. Differences of dynamical features
between the intracluster gas and member galaxies are independent of
cluster mass and concentration. We conclude that the intracluster gas
reaches the dynamic equilibrium state earlier than the almost 
collisionless member galaxies.
\end{abstract}


%%%%%%%%%%%%%%%%%%%%%%%%%%%%%%%%%%%%%%%%%%%%%%%%%%%%%%%%%%%%%%%%%%%%%%%
\section{Introduction}
\label{intro}
Galaxy clusters have been idealized to have spherically symmetric
structures in many studies, such as those on cluster mass estimation
\citep[e.g.,][]{app+10,pap+11}. Observations, however, indicate that
most clusters of galaxies have elliptical or irregular morphologies
\citep[e.g.,][]{me12}. In the hierarchical structure formation
scenario, galaxy clusters grow through mergers of smaller subclusters
or groups and the accretion of material \citep[e.g.,][]{bsb+09,mbb+09}. In
the early stage of merging, violent dynamic interactions occur between
subclusters, matter in the cluster is generally stretched
along the merging axis, the cluster therefore shows a bimodal or
elongated morphology \citep[e.g.,][]{rbl93,rbl96,rlb97}. After the
fast and violent merging stage, matter in the cluster roughly
follows a 3-dimensional Gaussian distribution and the cluster
shows an elliptical morphology \citep[e.g.,][]{l67}. After evolving
further for several Giga-years, the merged clusters will approach a
state of virial and hydrostatic equilibrium, and have almost
spherically symmetric structures \citep[see simulations,
  e.g.,][]{pfb+06}.

During cluster merging, the intracluster medium (ICM) evolves 
differently from the member galaxies and dark matter mainly
because the former is viscous while the latter are almost collisionless.
Observationally, the Bullet cluster \citep[1E 0657--56,
  e.g.,][]{mgd+02} is a textbook example: the distribution of the ICM
shows a large discrepancy to that of member galaxies and dark
matter \citep{cgm04,cbg+06}. Simulations have shown that the dark matter
of merging clusters generally exhibits a more elongated morphology
\citep{mog22} and requires a longer time to reach the equilibrium state
than the ICM \citep[see figure 15 and 16 in][]{pfb+06}. After reaching
the virilization state, relaxed clusters recover a good spatial 
matching between the ICM, member galaxies and dark
matter \citep[e.g.,][]{cs02,dem+11}. Due to continuous
radiative cooling of ICM, relaxed clusters are often highly
concentrated in X-ray with a very bright cool-core in the cluster
center \citep[see][as a review]{f12}.

Many efforts have been made to quantify the dynamical state of galaxy
clusters with optical or X-ray data
\citep[e.g.,][]{rbp+07,srt+08,wh13,yhw22}. The ellipticity has been
widely used to describe the global morphology of galaxy clusters. In
the optical, the cluster ellipticity can be estimated from the projected
distribution of member galaxies \citep[e.g.,][]{cm80,dkv95,sml+97}. In
X-rays, one can directly obtain the cluster ellipticity by fitting the
brightness image with a two-dimensional $\beta$-model
\citep[e.g.,][]{yh20,yw22}. \citet{bc92} studied five Abell clusters
and found that the ellipticities derived from the ICM are smaller than those
calculated by \citet{cm80} with member galaxies. \citet{sf94} reported
that more massive clusters tend to have a rounder morphology. The cluster ellipticity has weak or no redshift evolution
\citep[e.g.,][]{p02}.

Theories and simulations predict that the dynamic relaxation for the
almost-collisionless dark matter and member galaxies differs from that
of the diffuse ICM \citep[e.g.,][]{pfb+06}. Observational
differences between member galaxies and ICM have been established in
clusters in early stages of merging, such as in the Bullet cluster. However, does this
difference persist in clusters in later stages of dynamic evolution?
In this work, we report on systematical differences in dynamic
features between the distributions of member galaxies in the optical band
and the ICM expressed by the X-ray image. Recently, we processed the
X-ray images homogeneously for 1844 clusters with the archival data of
the {\it Chandra} and {\it XMM-Newton} satellites \citep{yh20,yhw22},
and therefore have a dataset available. In optical, a large
number of clusters have been identified with well-recognized member
galaxies \citep[e.g.,][]{whl12,wh15}. Therefore, it is now feasible to
study the distributions of member galaxies and the ICM for a large
sample of clusters. The distributions of the two components could be
very complex and vary among galaxy clusters in different dynamical
states. As the first step, in this paper we compare the dynamic
features of member galaxies and ICM only for clusters in a
quasi-equilibrium state.

\begin{figure*}
\centering
\includegraphics[width=0.31\textwidth, angle=-90]{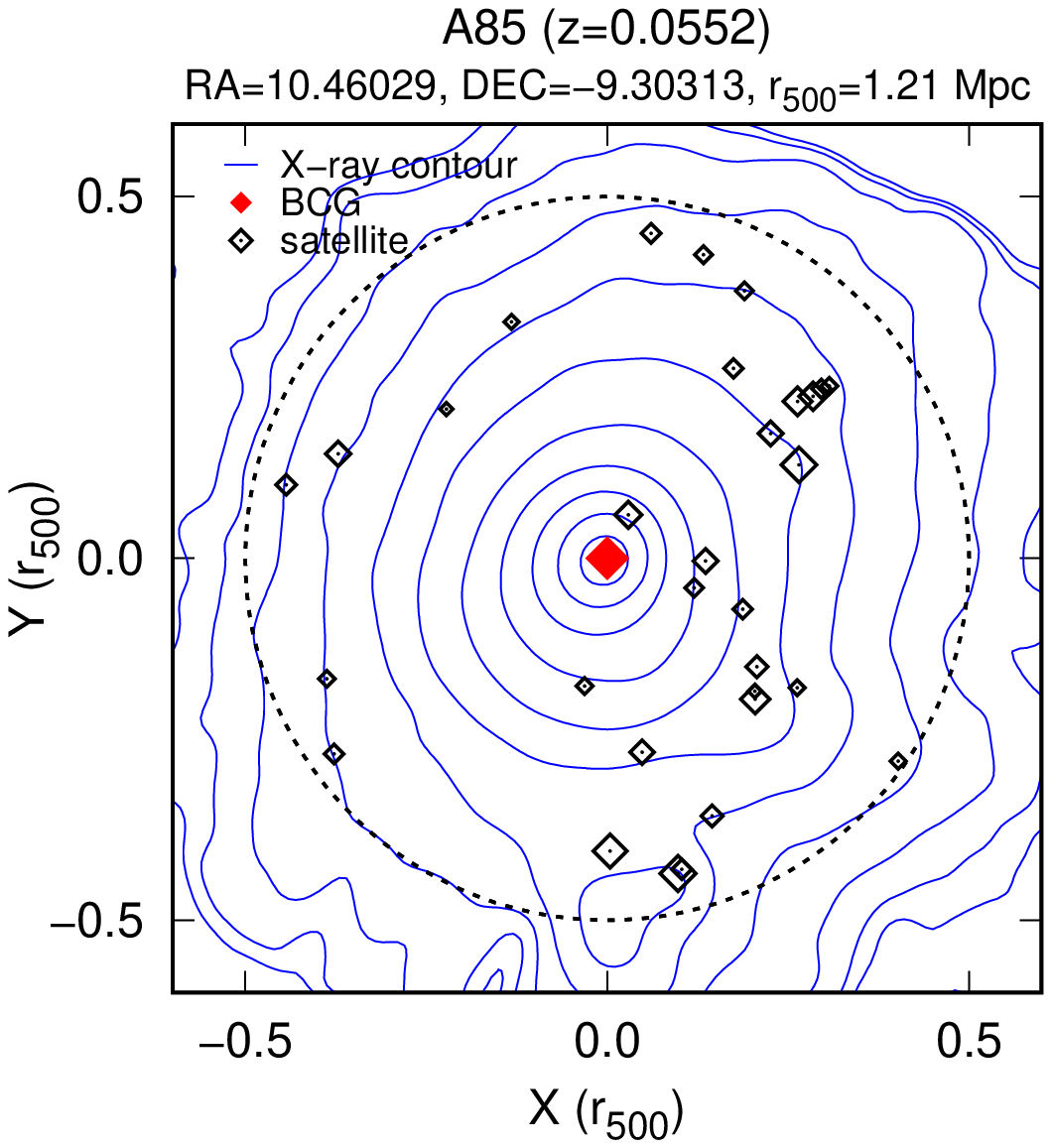}
\includegraphics[width=0.31\textwidth, angle=-90]{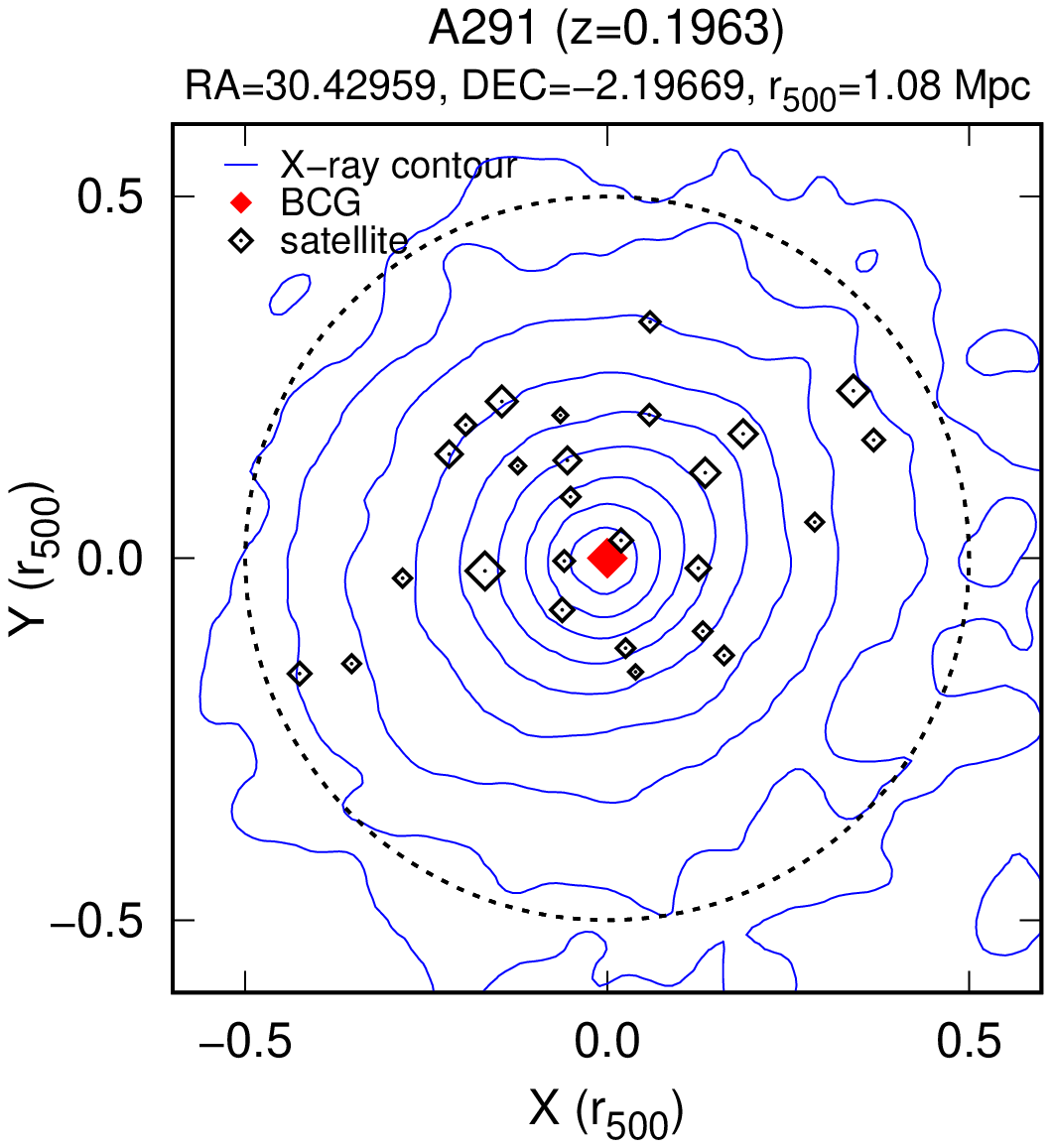}
\includegraphics[width=0.31\textwidth, angle=-90]{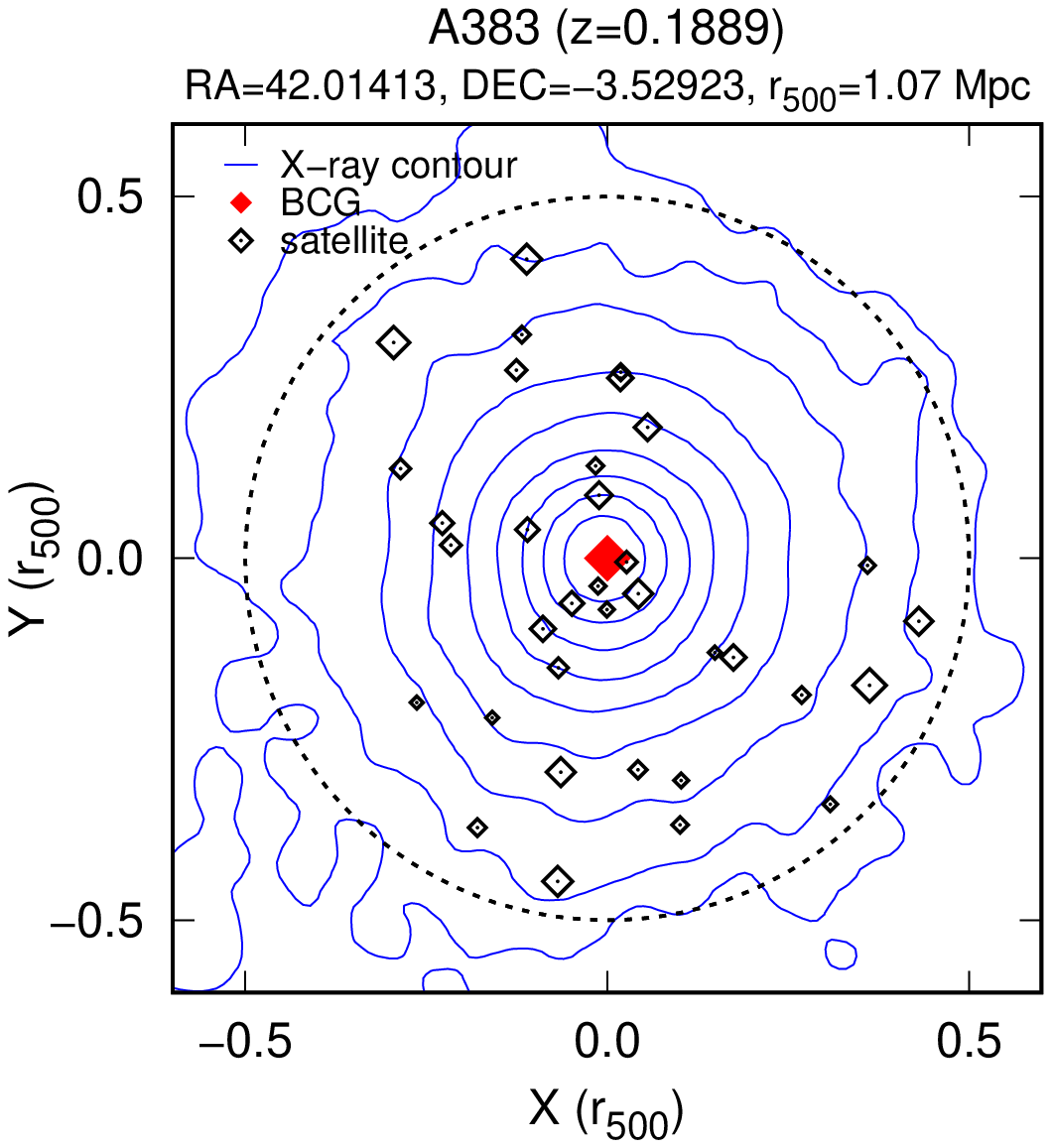}\\
\includegraphics[width=0.31\textwidth, angle=-90]{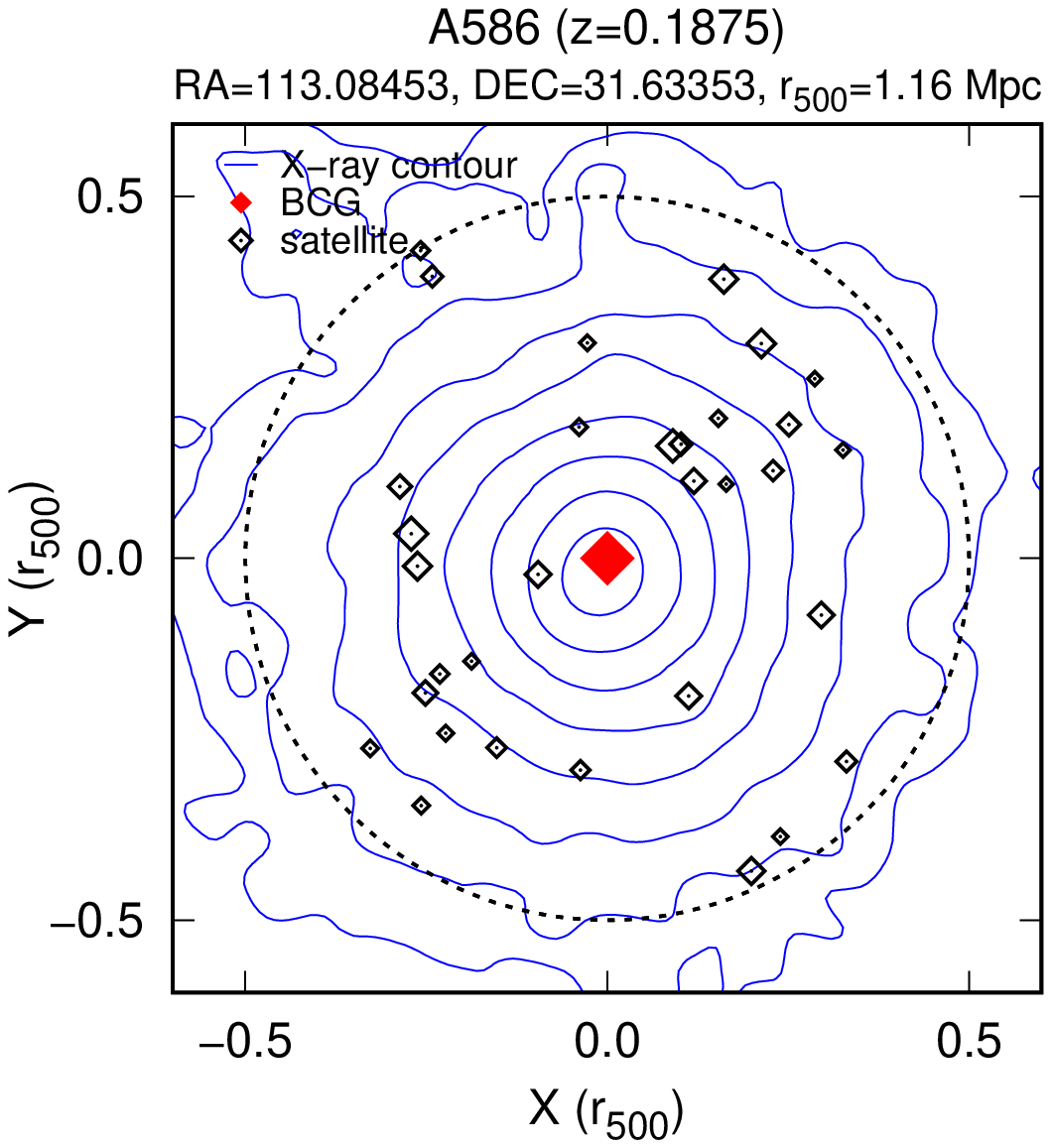}
\includegraphics[width=0.31\textwidth, angle=-90]{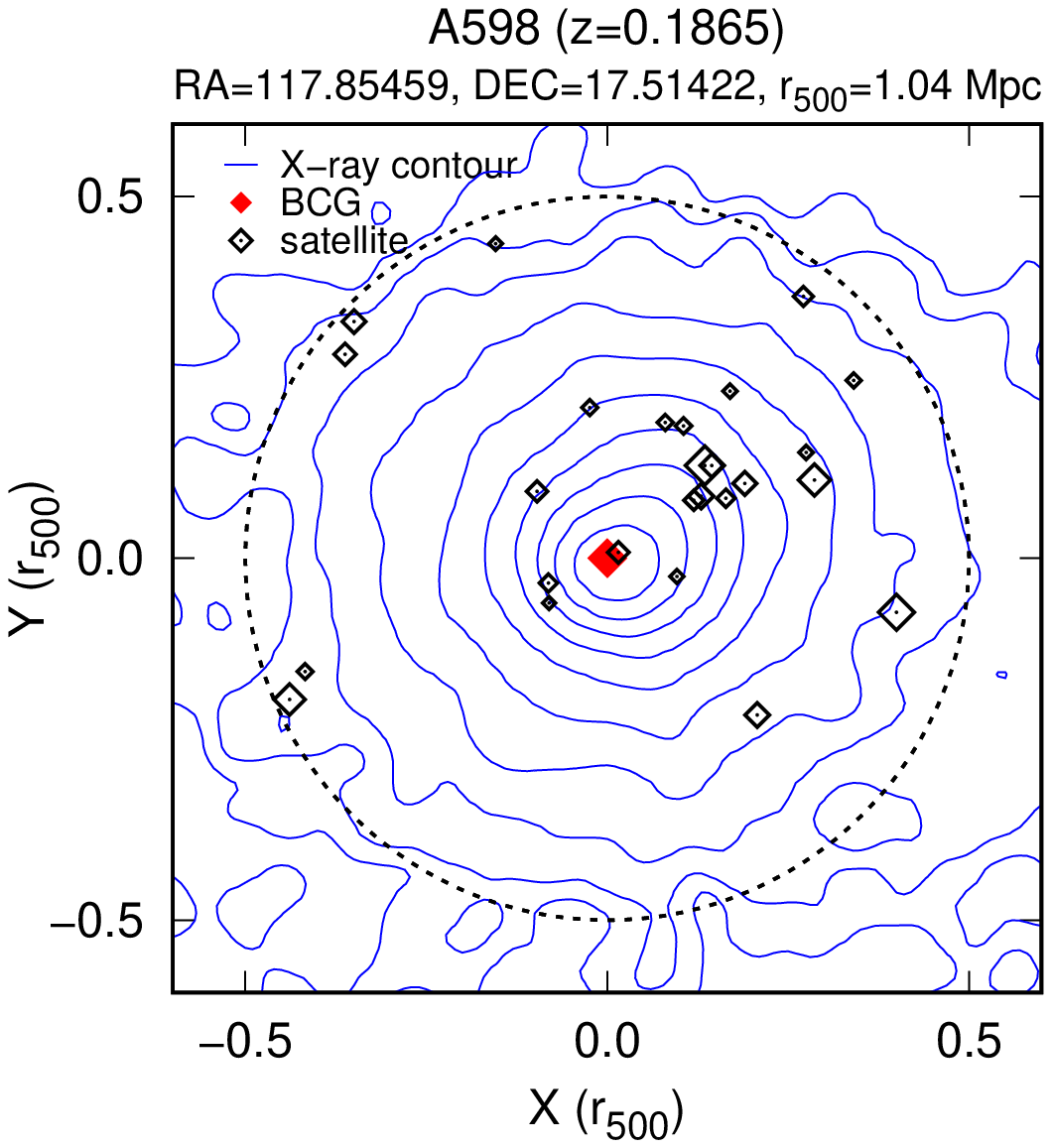}
\includegraphics[width=0.31\textwidth, angle=-90]{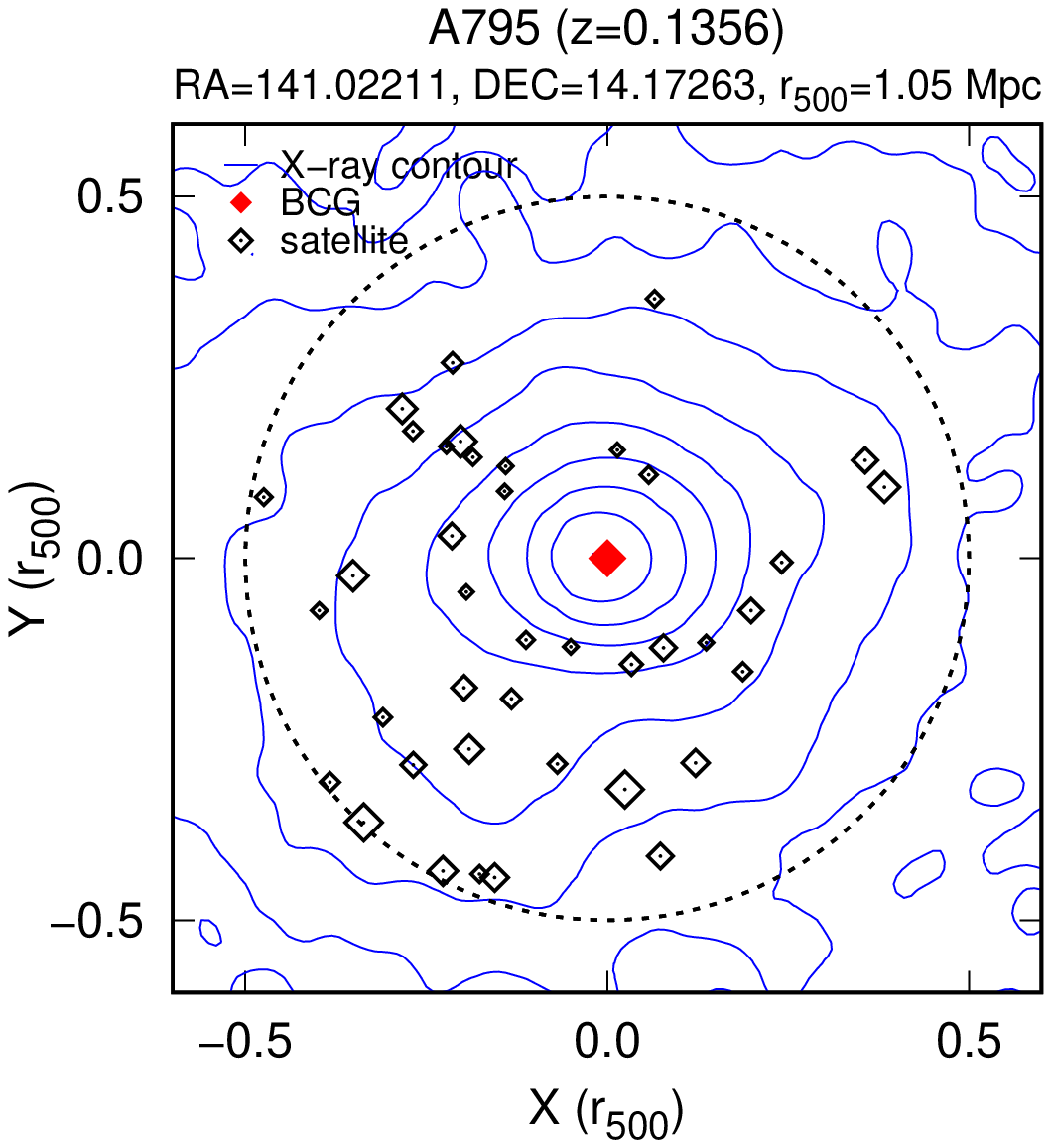}\\
\includegraphics[width=0.31\textwidth, angle=-90]{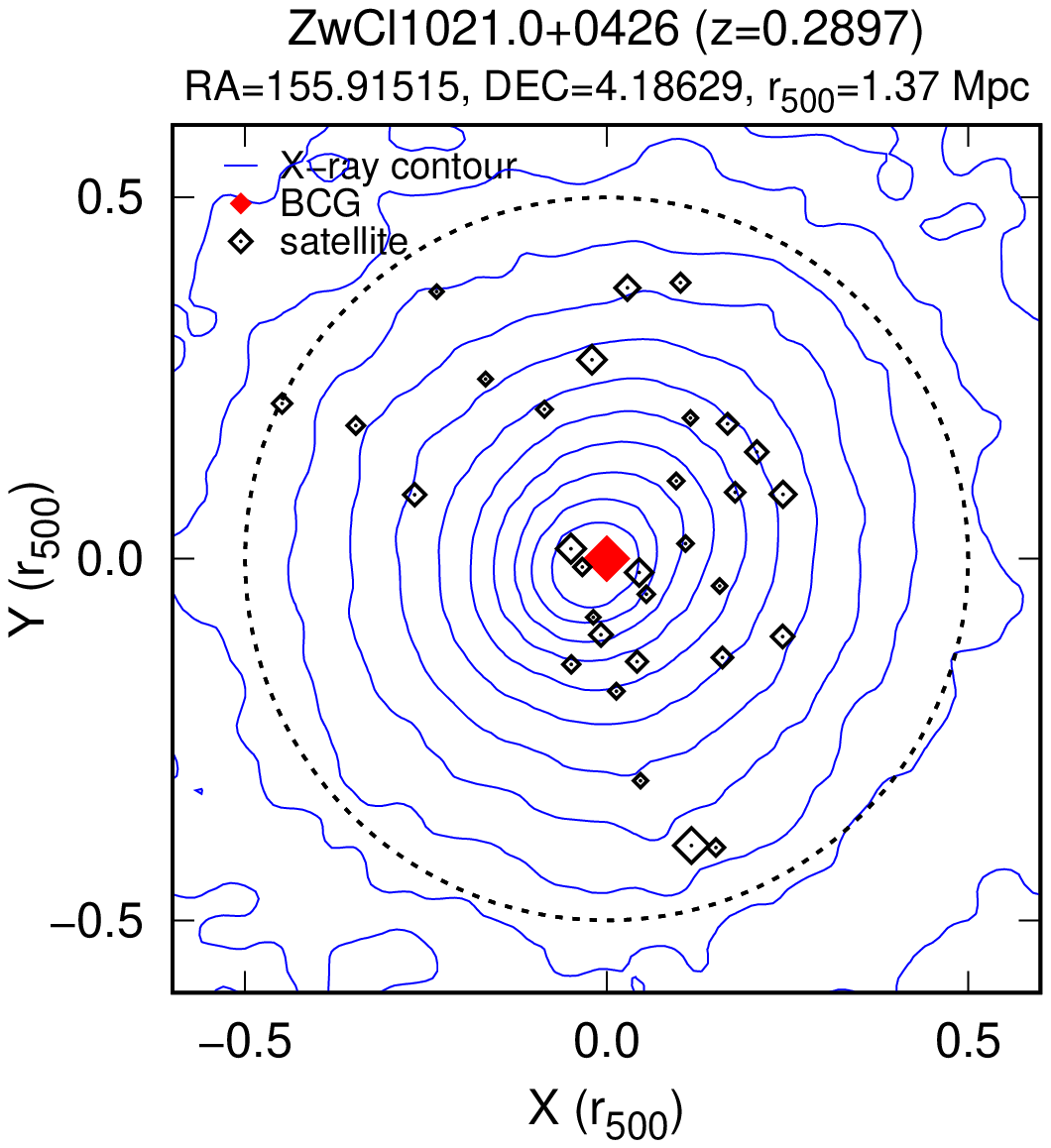}
\includegraphics[width=0.31\textwidth, angle=-90]{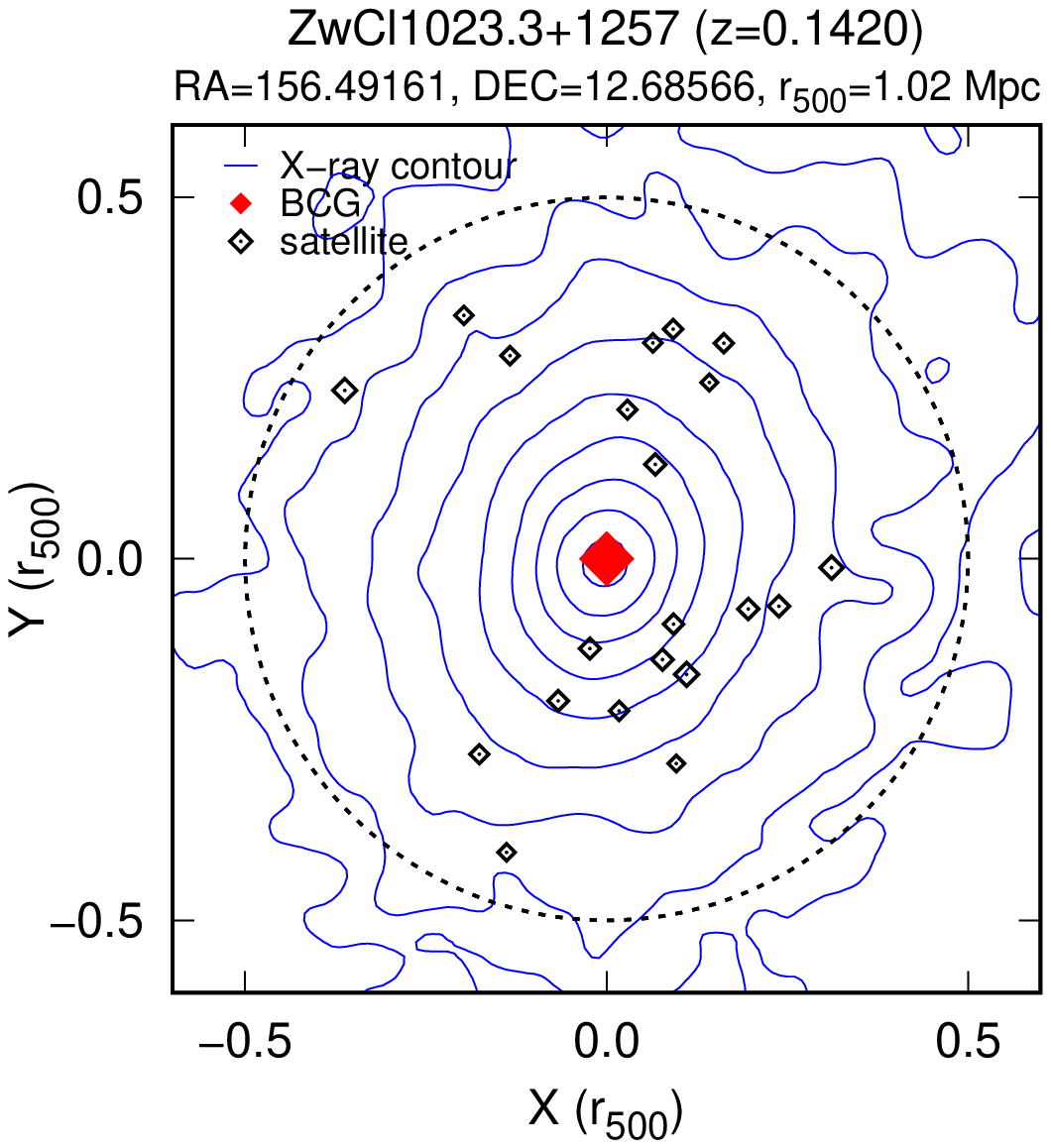}
\includegraphics[width=0.31\textwidth, angle=-90]{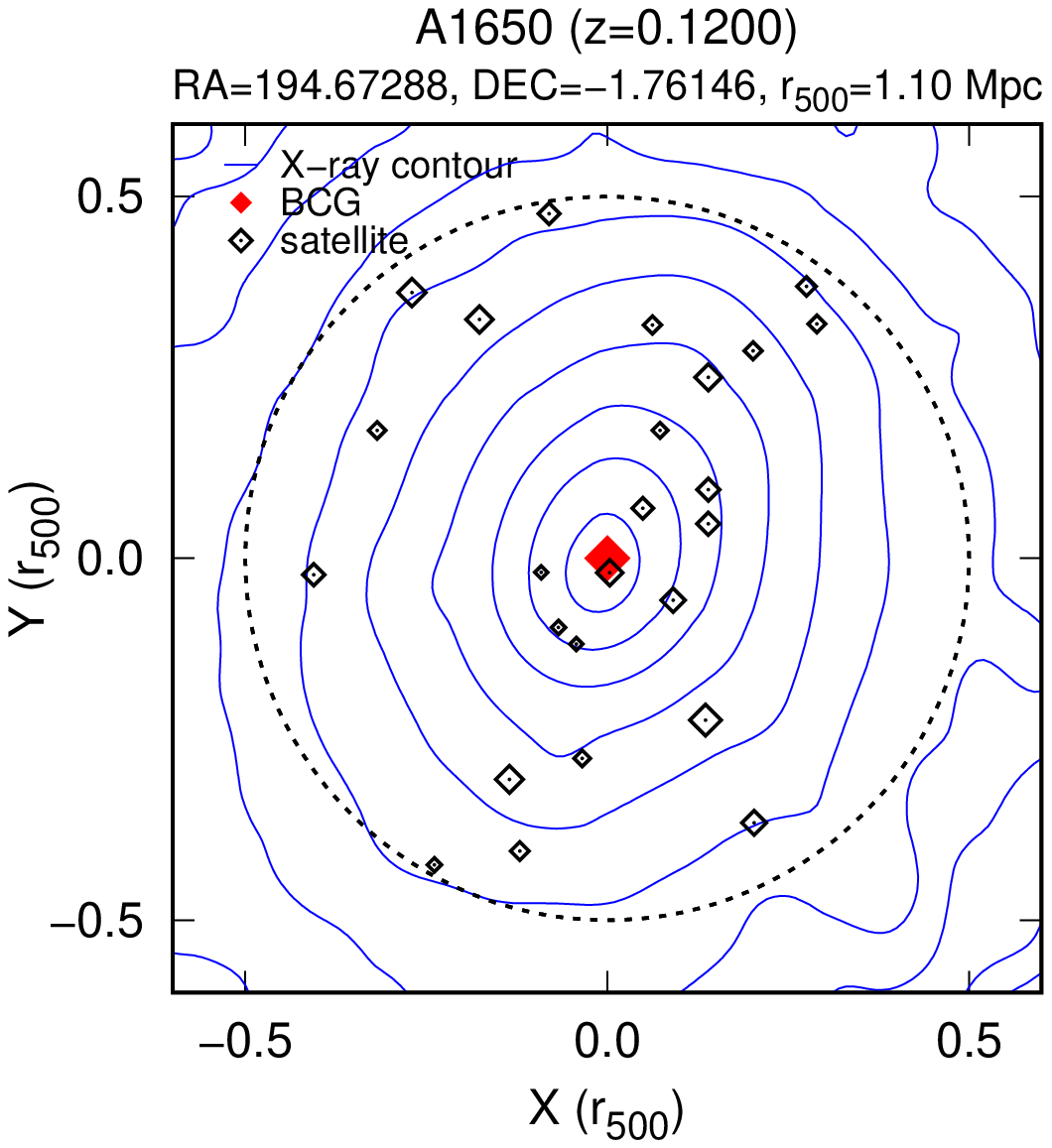}\\
\includegraphics[width=0.31\textwidth, angle=-90]{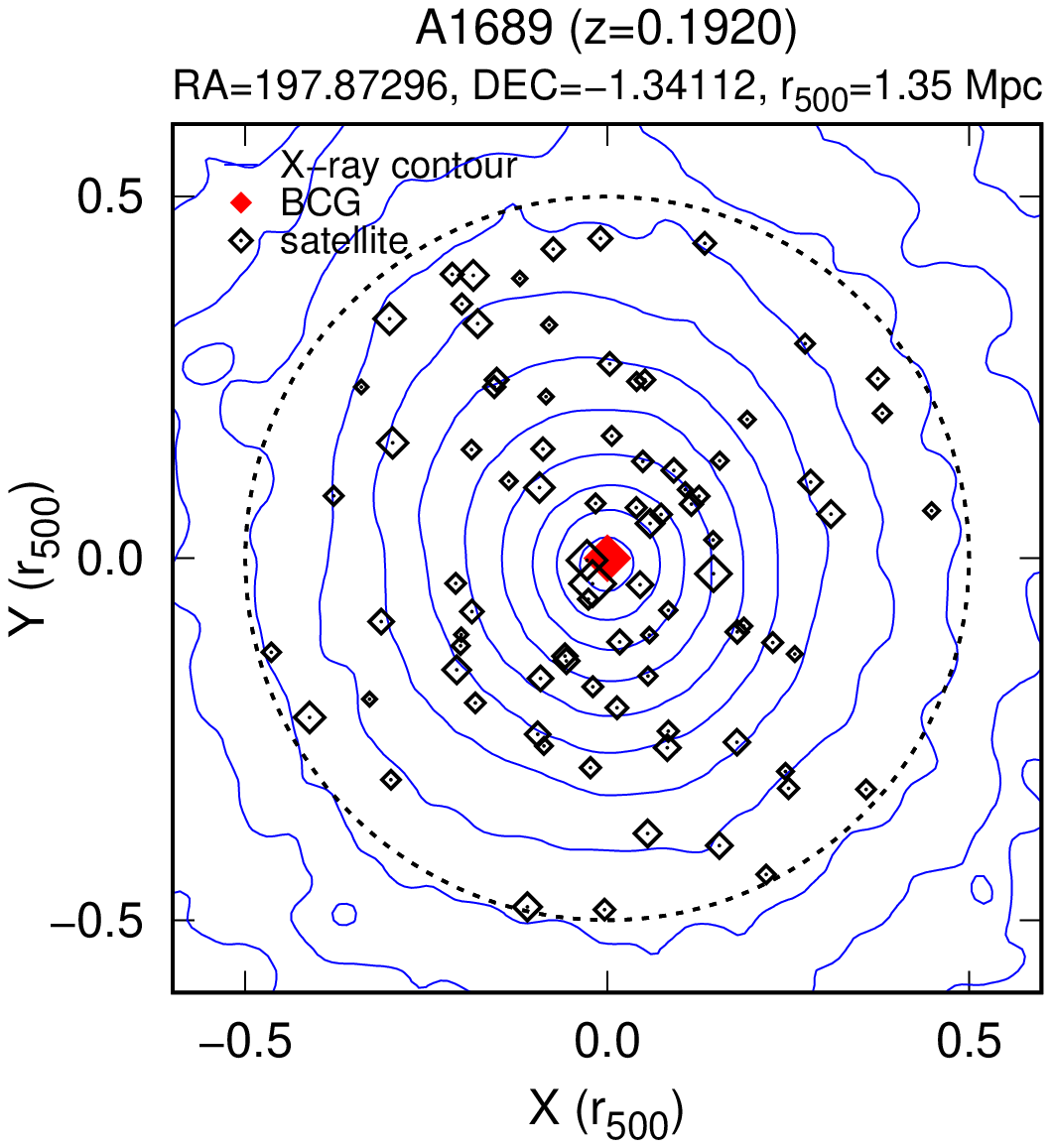}
\includegraphics[width=0.31\textwidth, angle=-90]{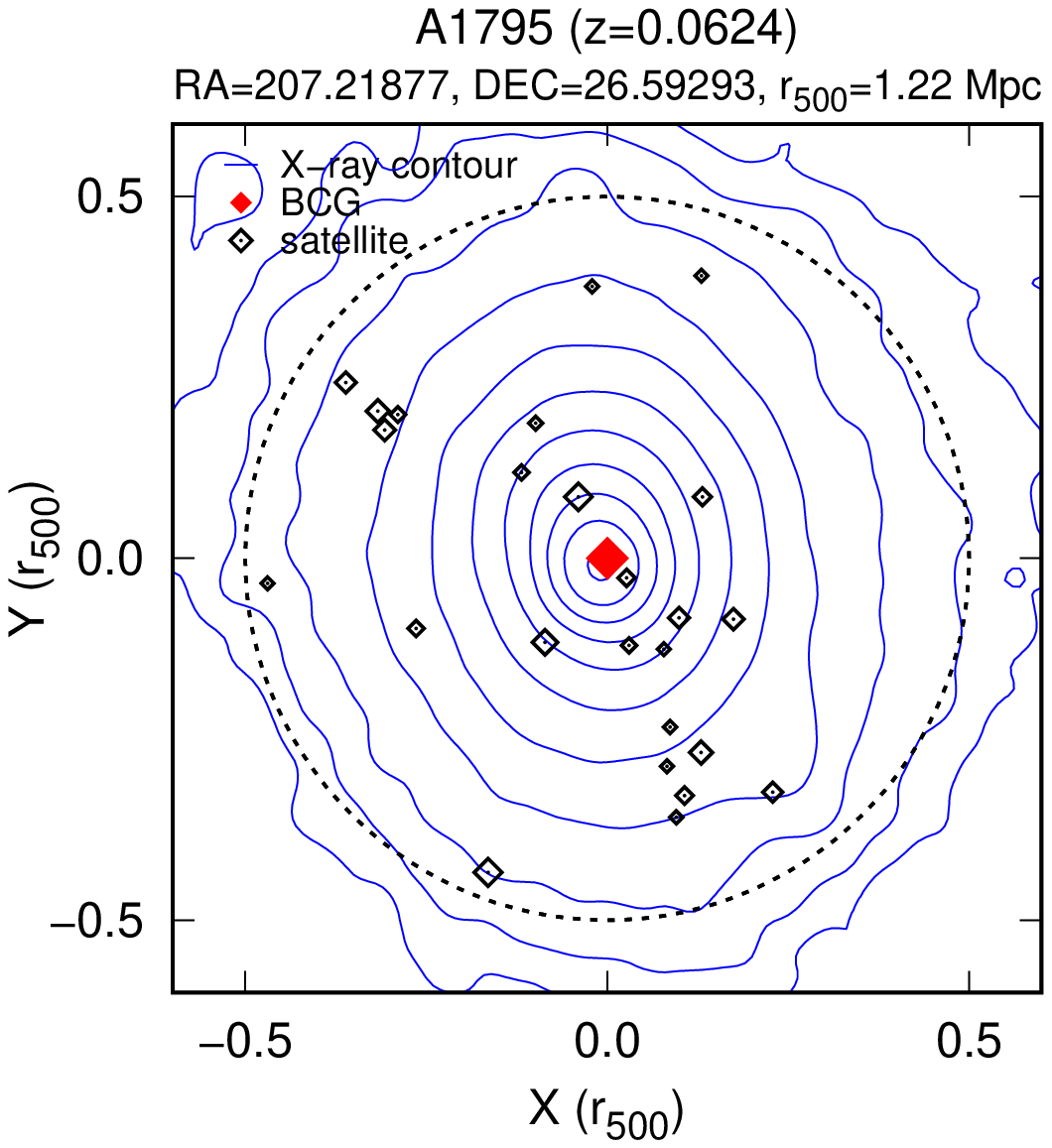}
\includegraphics[width=0.31\textwidth, angle=-90]{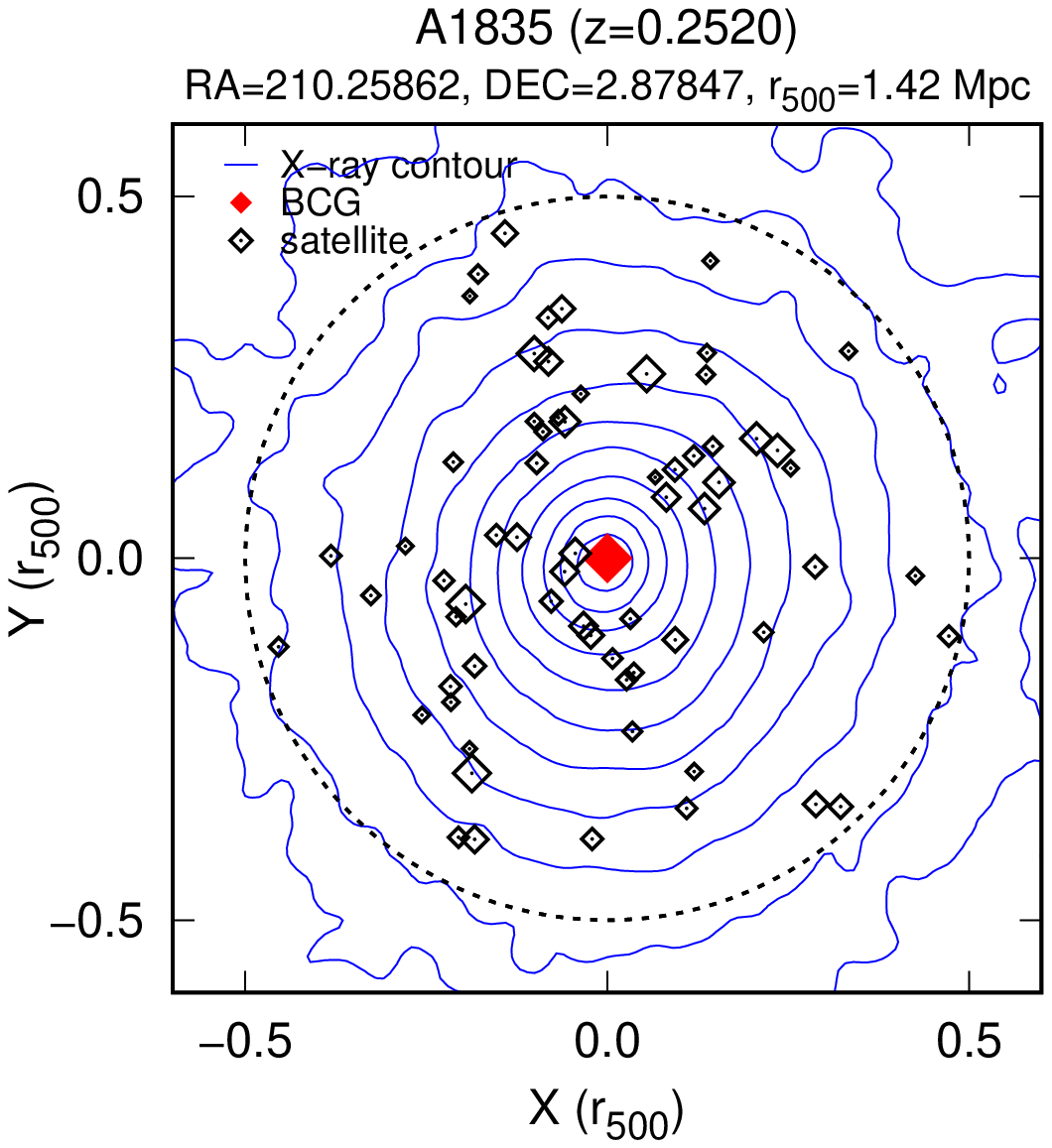}\\
\caption{- {\it continued}}
\end{figure*}
\addtocounter{figure}{-1}
\begin{figure*}
\centering
\includegraphics[width=0.31\textwidth, angle=-90]{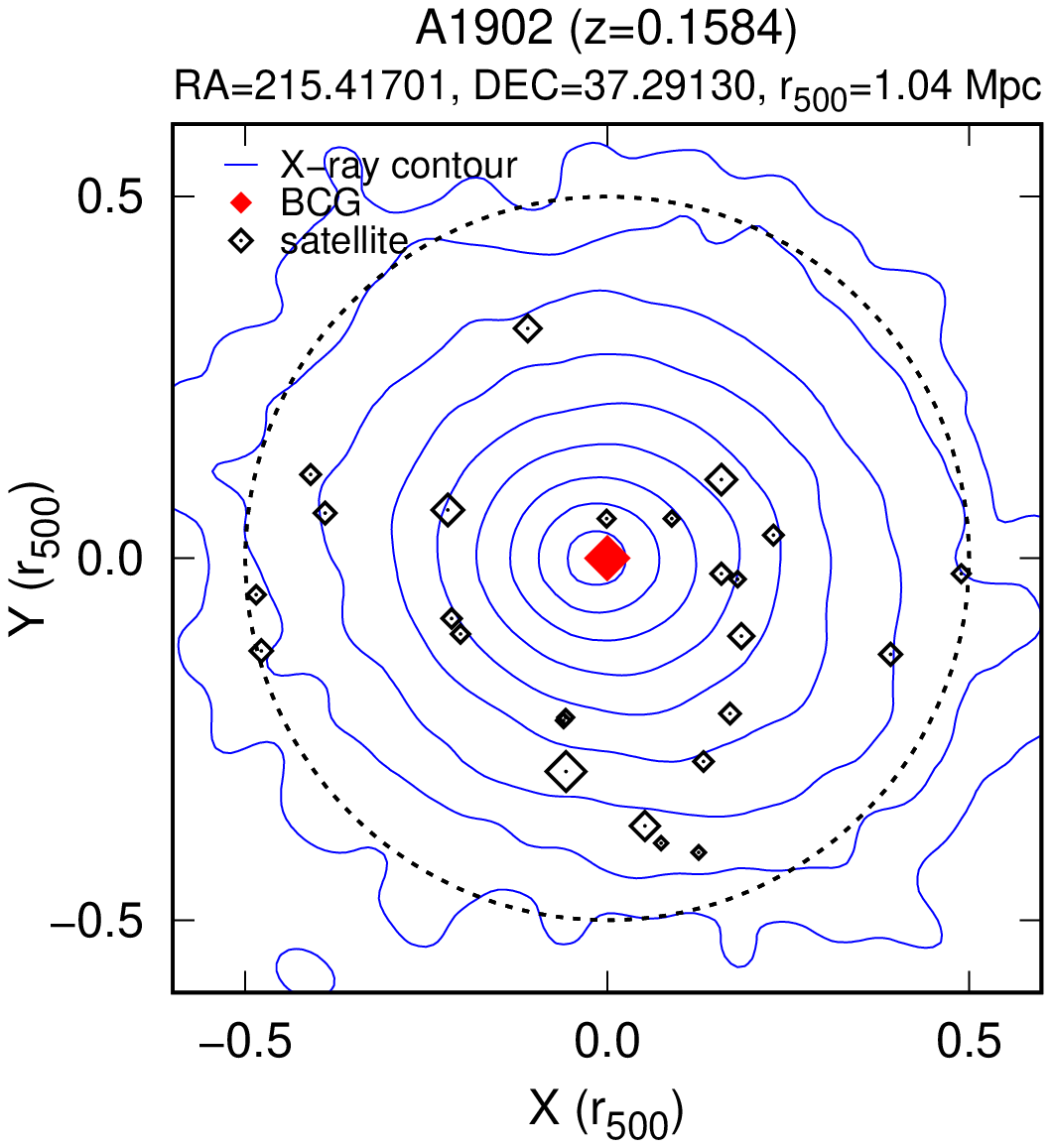}
\includegraphics[width=0.31\textwidth, angle=-90]{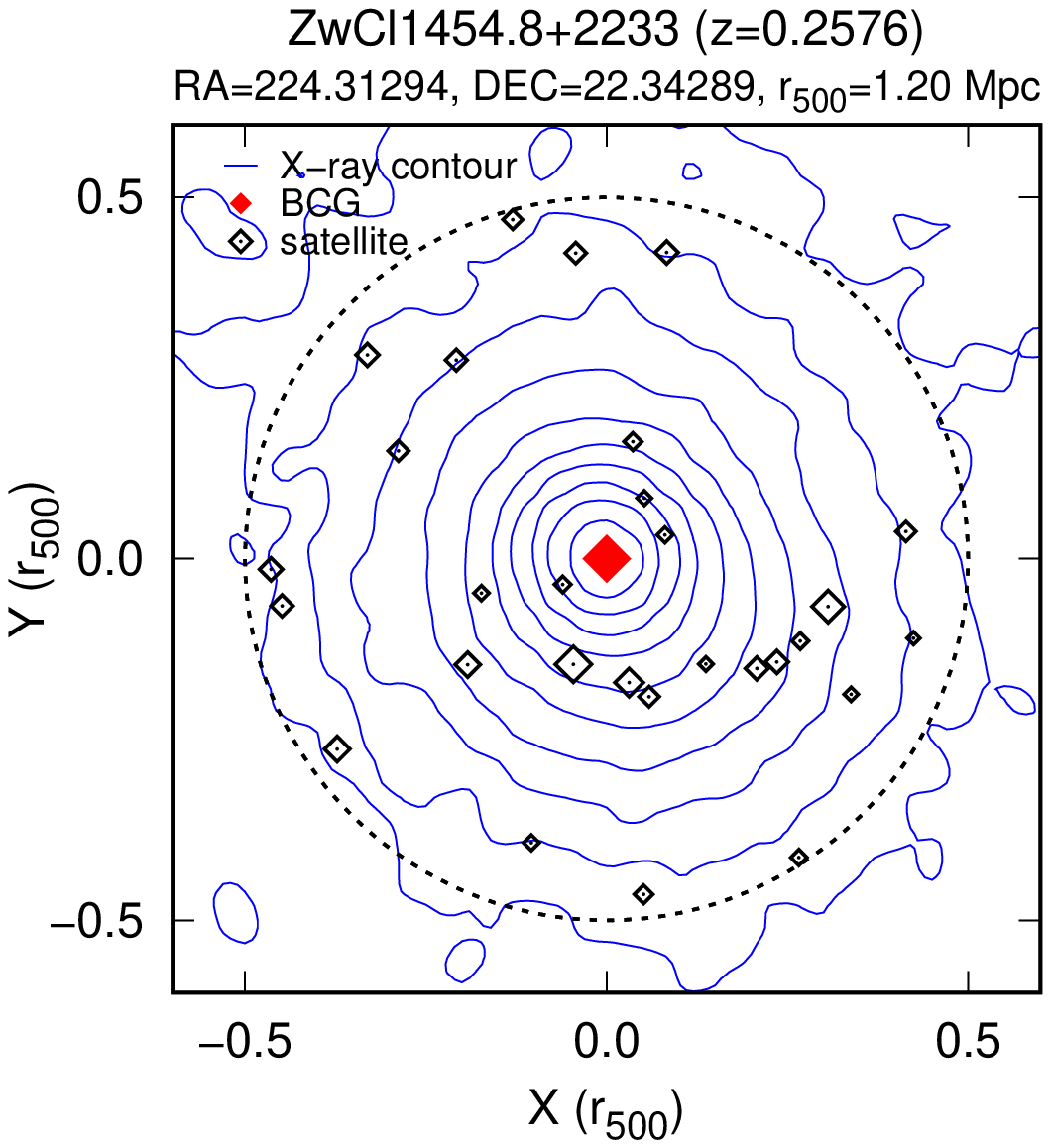}
\includegraphics[width=0.31\textwidth, angle=-90]{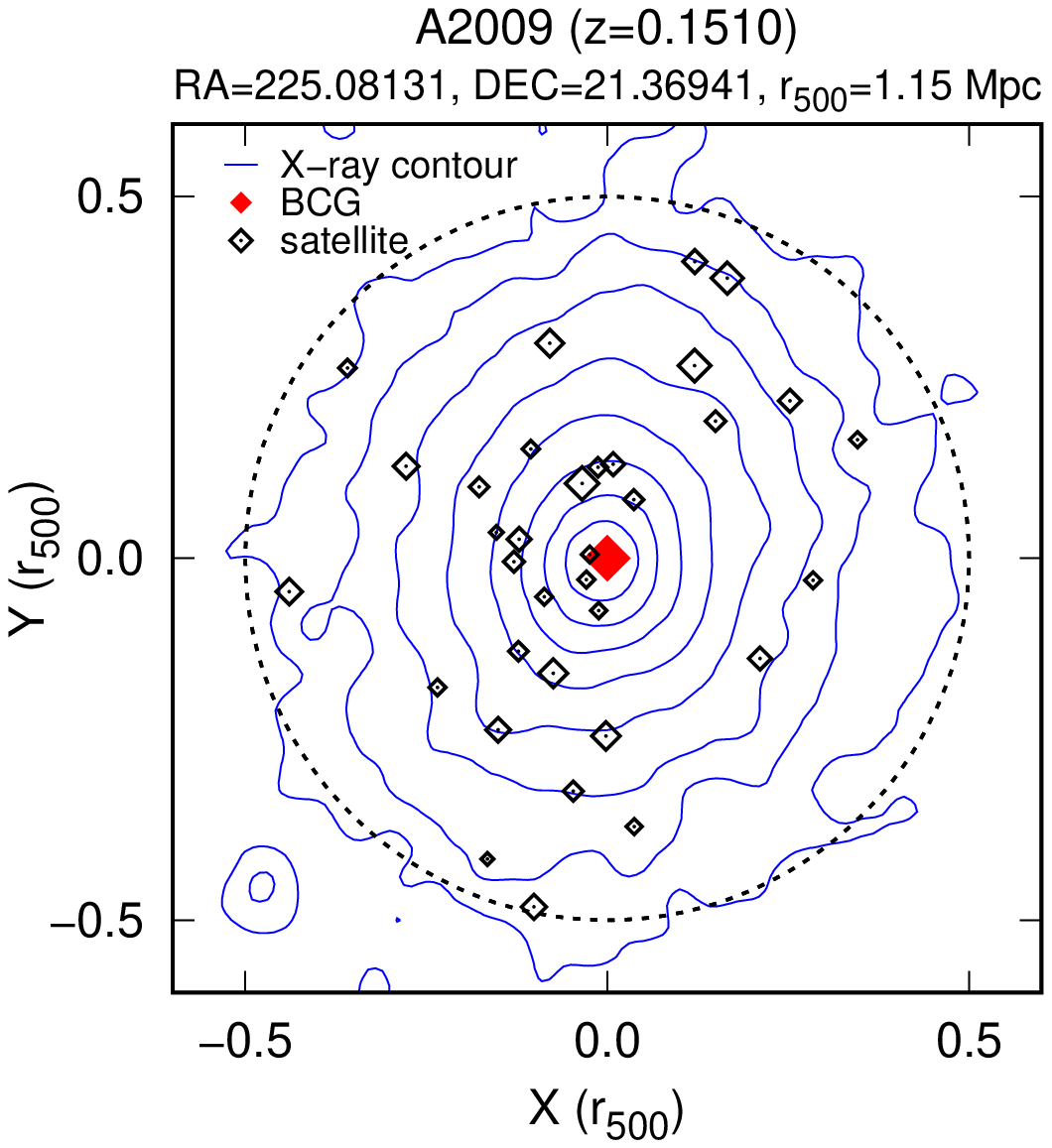}\\
\includegraphics[width=0.31\textwidth, angle=-90]{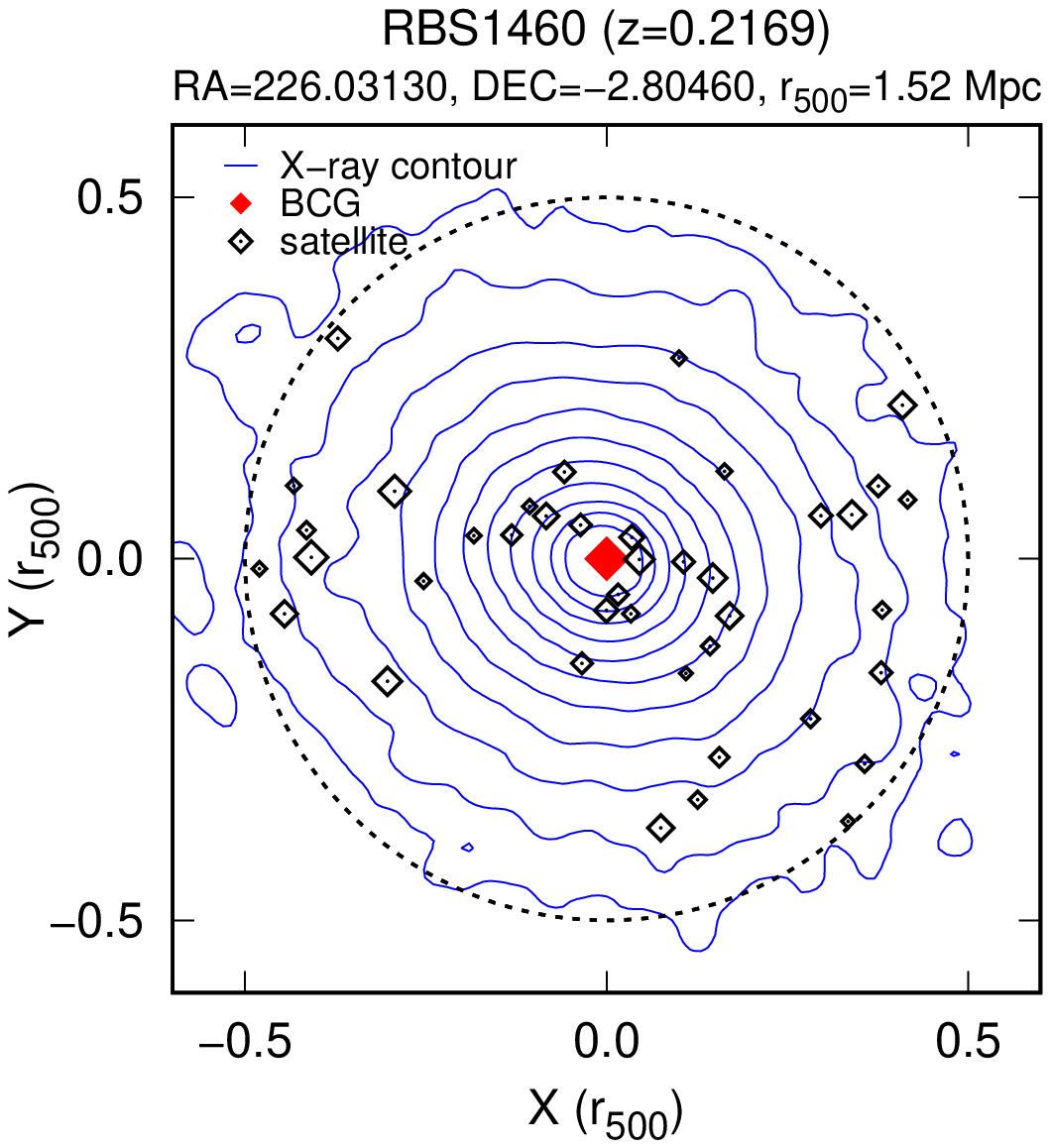}
\includegraphics[width=0.31\textwidth, angle=-90]{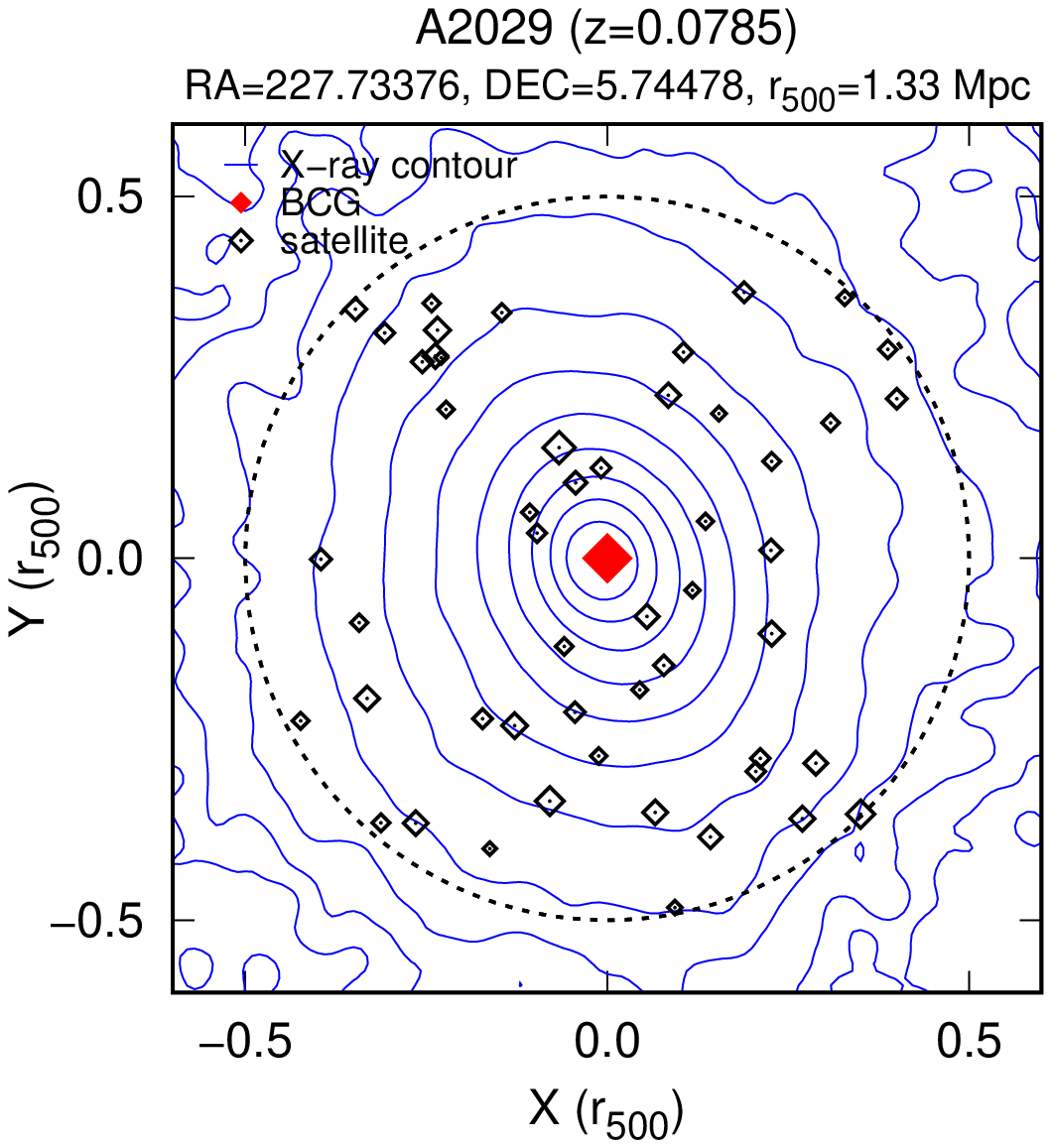}
\includegraphics[width=0.31\textwidth, angle=-90]{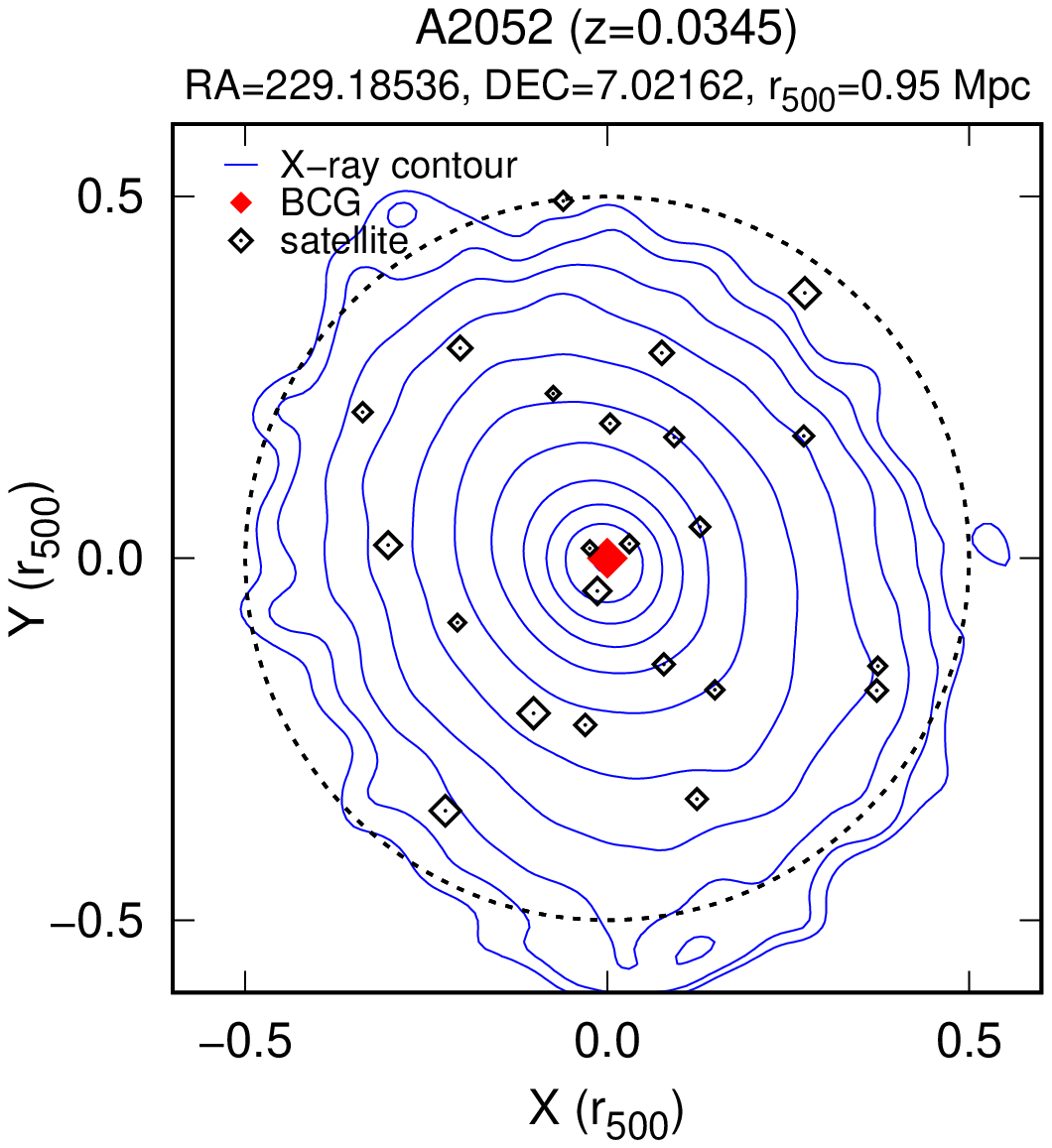}\\
\includegraphics[width=0.31\textwidth, angle=-90]{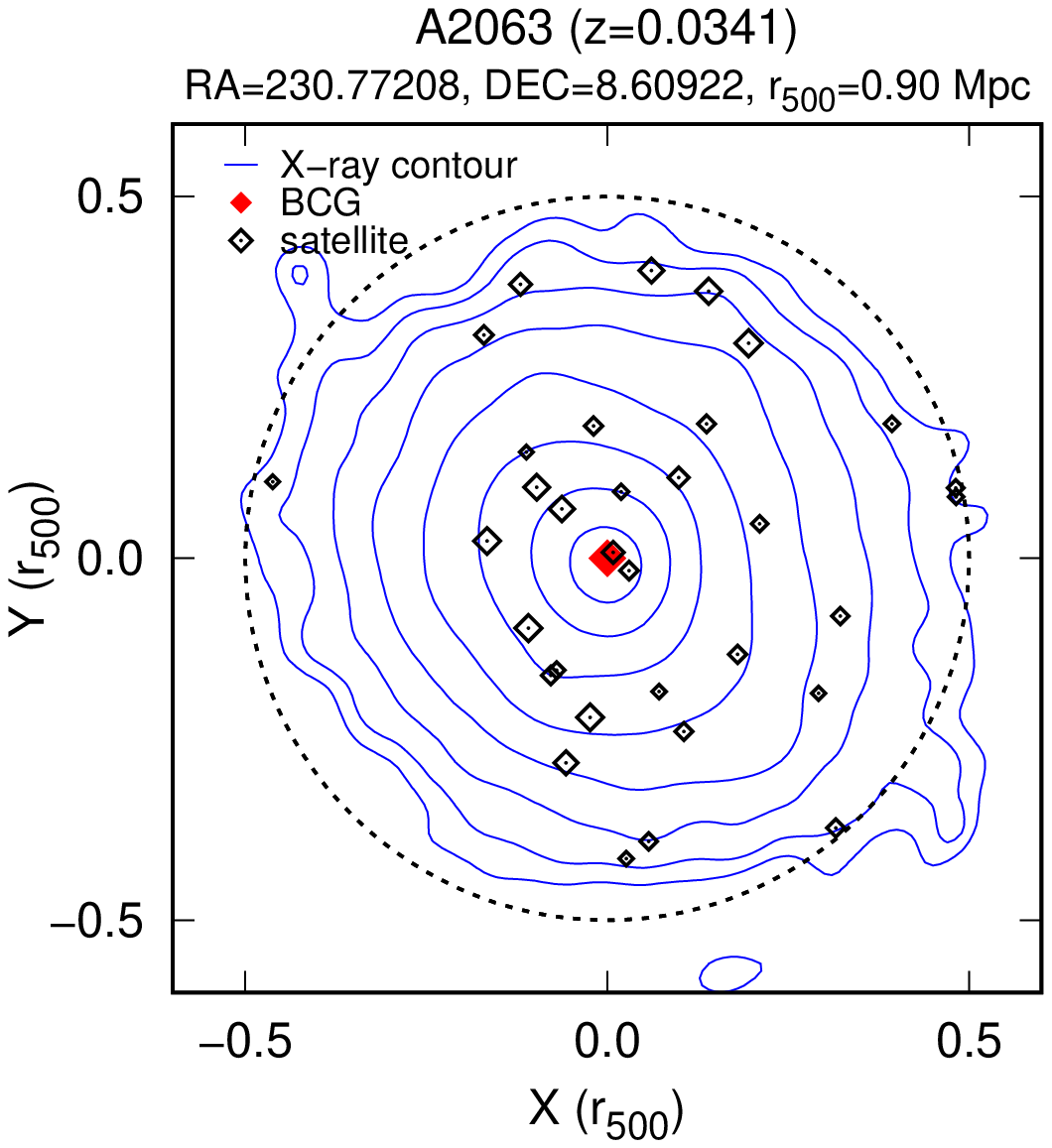}
\includegraphics[width=0.31\textwidth, angle=-90]{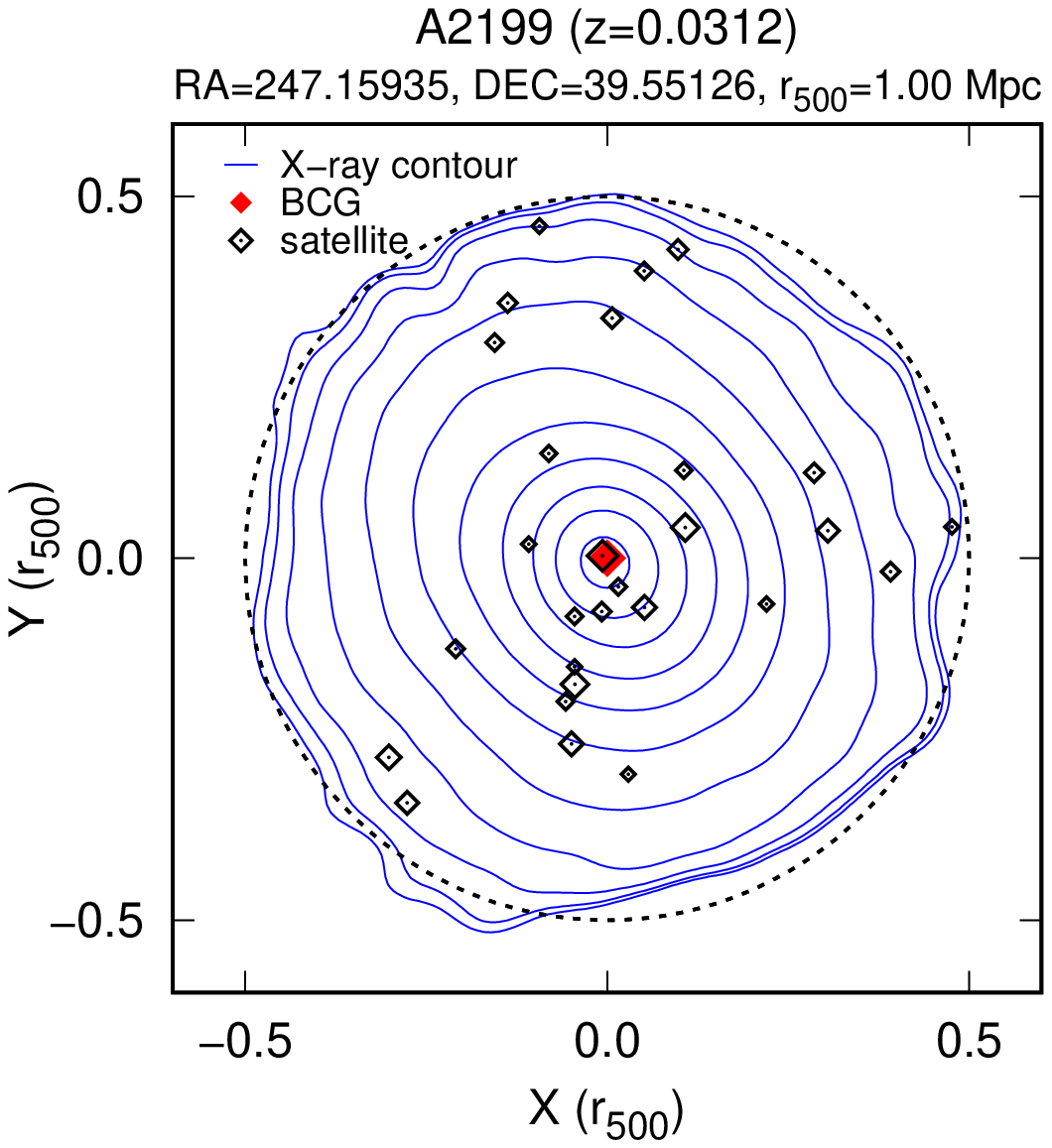}
\includegraphics[width=0.31\textwidth, angle=-90]{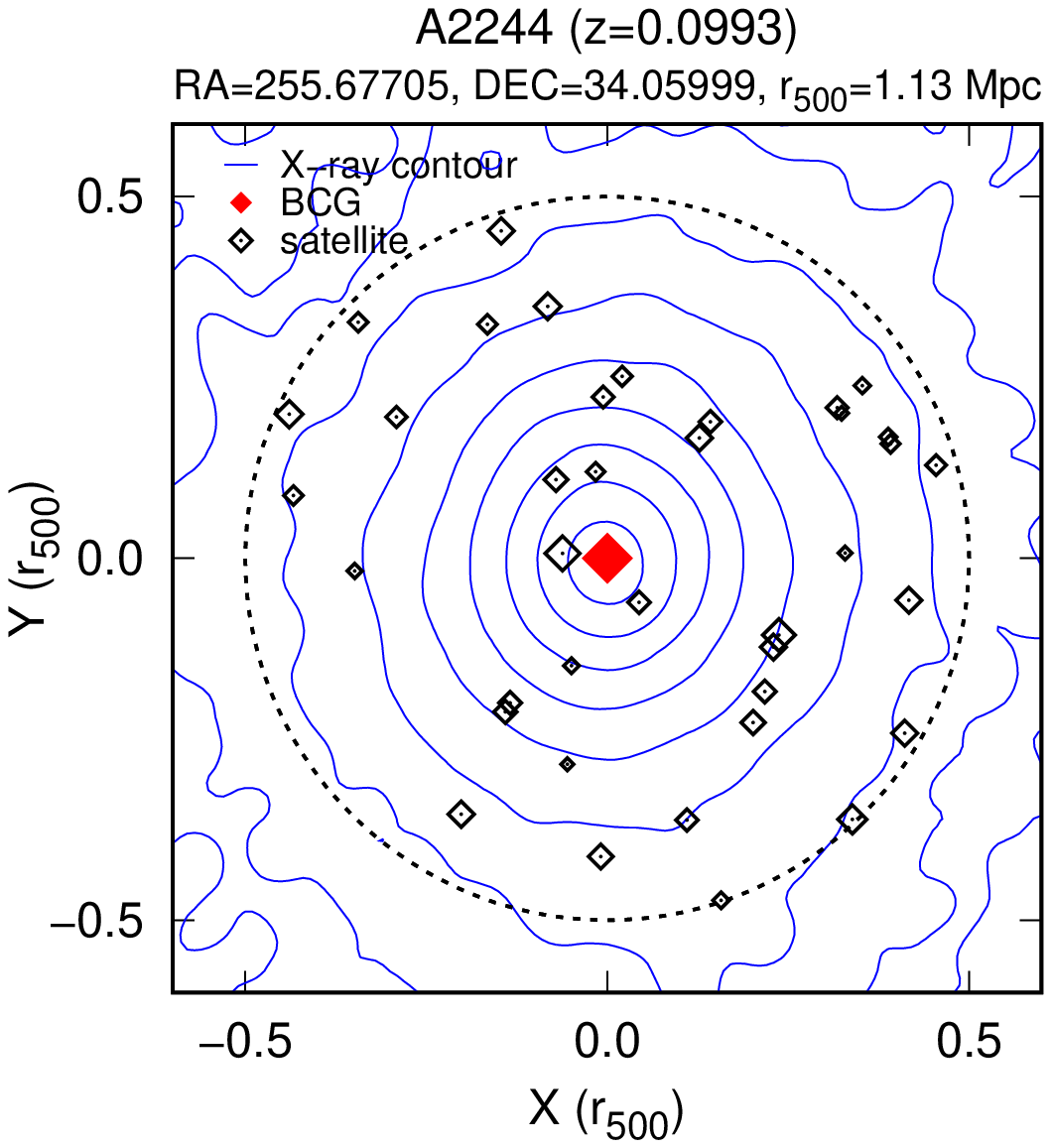}\\
\includegraphics[width=0.31\textwidth, angle=-90]{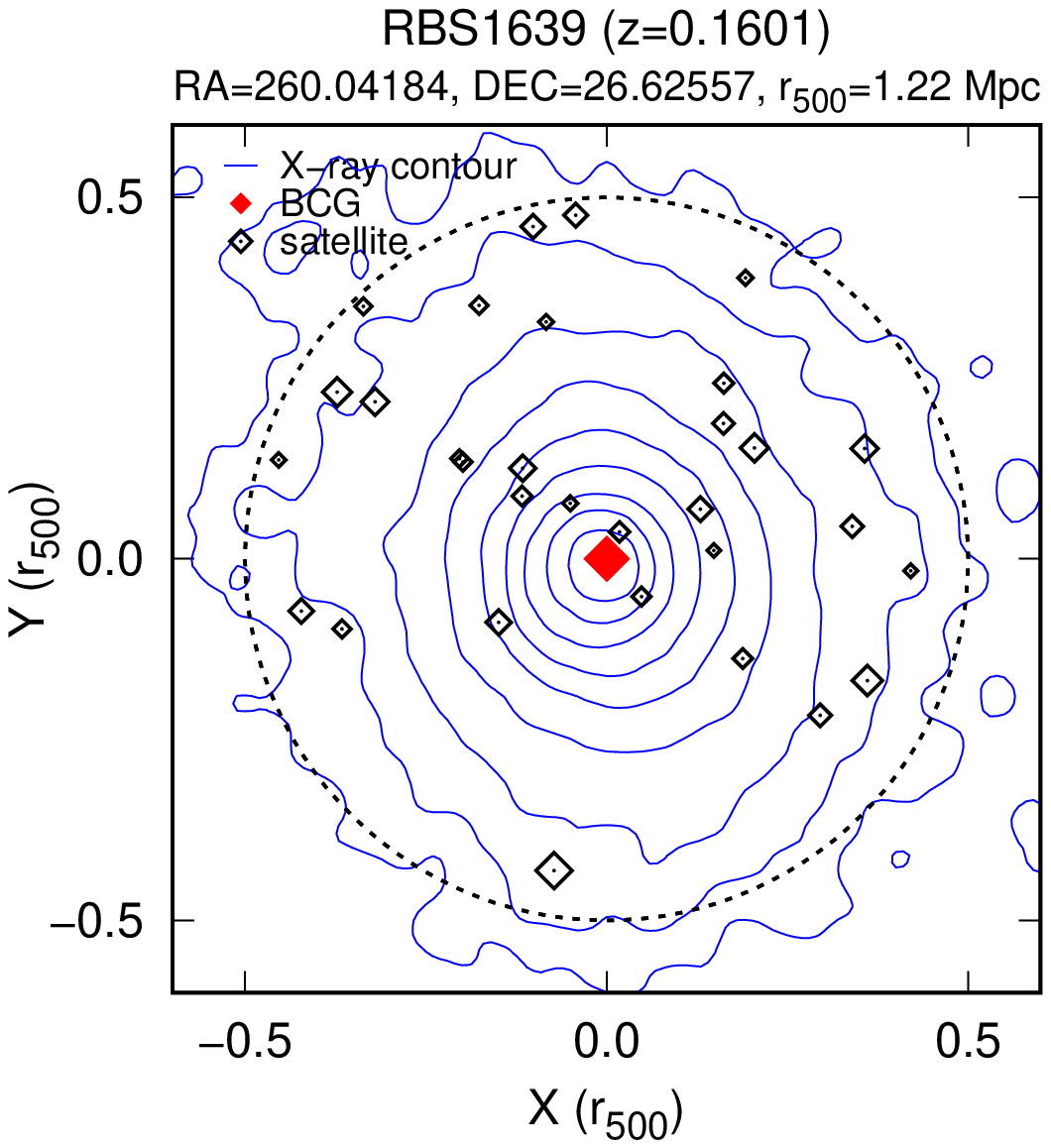}
\includegraphics[width=0.31\textwidth, angle=-90]{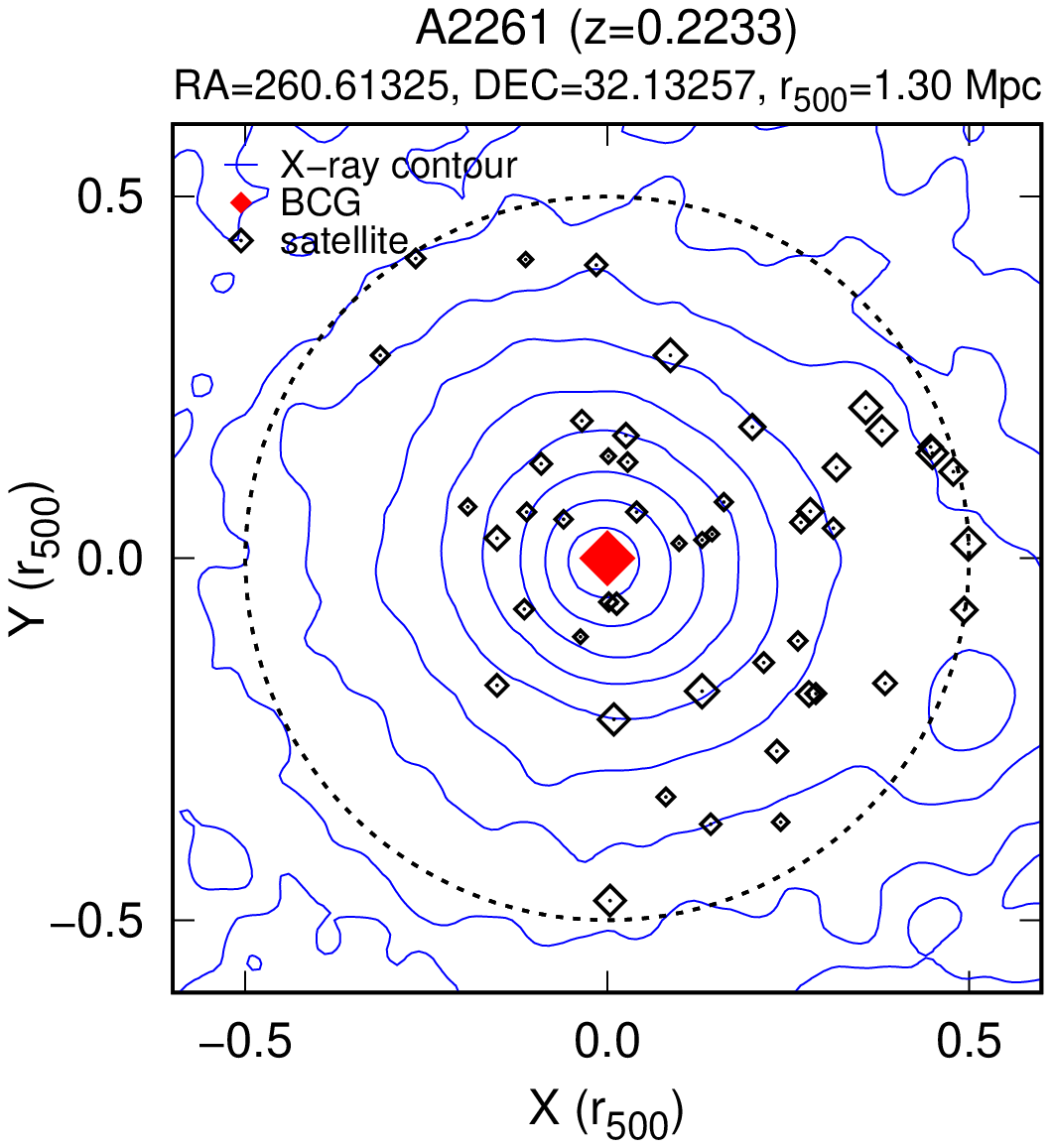}
\includegraphics[width=0.31\textwidth, angle=-90]{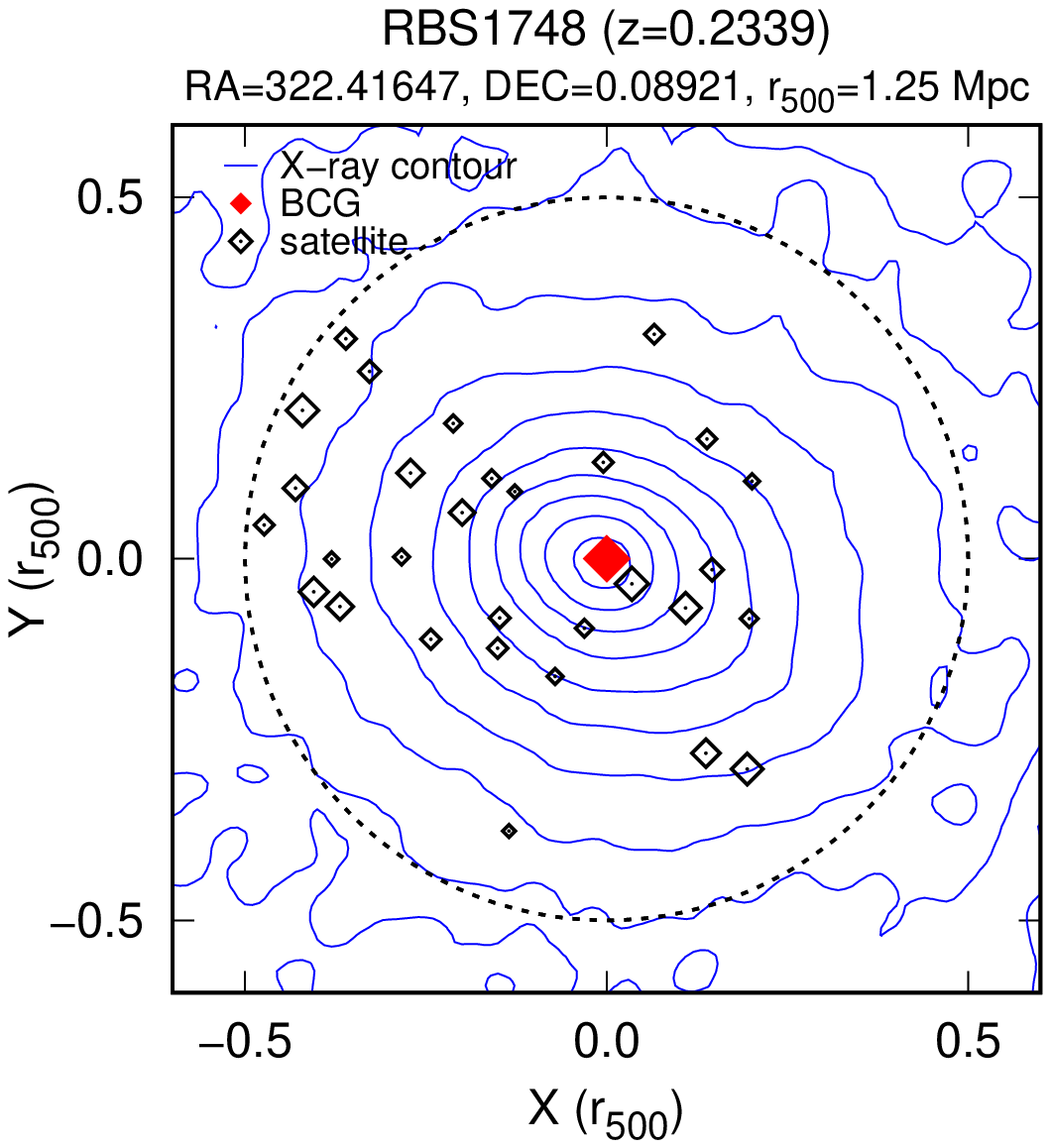}
\caption{X-ray images with superimposed member galaxies for the 24 clusters. The
  X-axis and Y-axis of each image are scaled with $r_{500}$. The
  coordinate, redshift and physical scale of $r_{500}$ for each
  cluster are written on the top of each panel. Contours indicate the
  distribution of X-ray brightness, at levels of $\langle S_{\rm
    bg}\rangle+3\times2^{n}\sigma$ cnt s$^{-1}$ deg$^{-2}$. Here
  $\langle S_{\rm bg}\rangle$ and $\sigma$ are the mean and
  fluctuation of the background, which are calculated in a clean
  region around the cluster, and $n=0, 1, 2, ...$. The solid and open
  diamonds denote the BCG and satellite galaxies respectively, and their sizes
  correspond to the luminosity of the galaxy. The large dotted circle shows
  the radius of $0.5r_{500}$, within which both the
  optical and X-ray parameters are calculated.}
\label{fig1}
\end{figure*}

The paper is structured as follows. In Section 2, we describe the methods
we used to select the cluster sample and derive the parameters.
In Section 3, we discuss the dynamic features of member galaxies and
ICM. A brief summary is given in Section
4. Throughout this paper, a flat $\Lambda$CDM cosmology is adopted
with $H_0=70$ km~s$^{-1}$ Mpc$^{-1}$, $\Omega_m=0.3$ and
$\Omega_{\Lambda}=0.7$.

\section{Cluster sample and parameters}
\label{data}
\label{sample}
By processing the archival data of the {\it Chandra} and {\it XMM-Newton}
satellites, we obtained X-ray images for 1844 galaxy clusters and
calculated four kinds of dynamical parameters for
them \citep{yh20,yhw22}. The clusters are collected from both targeted
and serendipitous observations. The data of the two satellites are
processed with similar strategies: only photons in 0.5--5 keV are used;
the point sources are carefully identified and replaced by the
brightness of their ambient regions; the images are exposure corrected
and background subtracted; the X-ray images are smoothed to
a physical scale of 30 kpc to avoid an observational bias that the pixel size 
indicates different physical scale for clusters at different redshifts. 
Detailed information on data processing can be
found in the Section 2 of \citet{yh20} and \citet{yhw22}.

The optical data for member galaxies of clusters are taken from the
{\it Sloan Digital Sky Survey (SDSS)} following
\citet{wh15}. Galaxies are recognized as cluster members when their
$r$-band evolution-corrected absolute magnitude $M^{\rm e}_{\rm r}$
are brighter than -20.0 mag and their velocity differences from the
cluster are less than 2500 km s$^{-1}$ if spectroscopic redshifts 
are available or the redshift differences are less than $0.04(1+z)$ 
if only photometric redshifts are available \citep{wh15}. \citet{whl09} 
showed that using only photometric redshifts yields a completeness  about 90\% and a contamination rate of about 20\% for member
galaxies when $z<0.42$. 

We select a sample of galaxy clusters for this work with the following
four criteria. (1) The cluster is in quasi-equilibrium when ${\rm
  log_{10}}(c)>-0.5$, where $c$ is the concentration index. Here we
use the concentration index $c$ instead of the morphology
index $\delta$ because the latter is defined from the cluster
ellipticity \citep{yh20}. (2) The X-ray image of the cluster has a
good quality within 0.5$r_{500}$, i.e., the region encircled by
0.5$r_{500}$ is well covered by the 3$\sigma$ contour of the X-ray
image. Here $r_{500}$ is the radius in which the mean matter density
of the cluster is 500 times of the local critical density. (3) The
cluster redshift should be lower than 0.3 to ensure that member
galaxies with $r$-band absolute magnitude brighter than -20.0 mag are highly
complete ($>90\%$). We visually checked the optical images of these clusters to avoid contamination of foreground stars.
(4) The number of satellite galaxies in 0.5$r_{500}$, $N_{\rm sat}$,
is larger than 20 to ensure that the calculated parameters are
reliable. Finally, we get 24 clusters satisfying all the above
criteria, see Table~\ref{tab1}. The X-ray images with superimposed
distribution of member galaxies for the 24 clusters are shown in
Fig.~\ref{fig1}.

We collect various parameters of the clusters from literature. The
coordinate and redshift of the BCG measured from the {\it SDSS} data, see \citet{whl12} for details, are
regarded as that of their host clusters. Clusters with $z<0.05$ are
not contained in the catalogue of \citet{wh15}, we thus identify the
BCG and other member galaxies with the same method for A2052
($z=0.0345$), A2063 ($z=0.0341$) and A2199 ($z=0.0312$). The
concentration indices $c$ are taken from \citet{yhw22}. The cluster
radius $r_{500}$ and mass $M_{500}$ come from the catalogue compiled
by \citet{pap+11}. 

Virialized galaxy clusters generally show a small offset between the
brightness peak and the flux-weighted centroid, but the offset could be
large for disturbed clusters. We define a vector named the normalized
centroid offset $\vec{\eta}$ as
\begin{equation}
\label{eqt}
 \vec{\eta}=\frac{\vec{\mu}_{10}+\vec{\mu}_{01}}{r_{500}},
\end{equation}
where
\begin{equation}
\begin{split}
& \vec{\mu}_{10}=\frac{1}{\sum\limits_{i}f_{{\rm w},i}}{\sum\limits_{i}f_{{\rm w},i}}\cdot \vec{x}_{i} ,
& \vec{\mu}_{01}=\frac{1}{\sum\limits_{i}f_{{\rm w},i}}{\sum\limits_{i}f_{{\rm w},i}}\cdot \vec{y}_{i} .
\end{split}
\end{equation}
The $f_{{\rm w},i}$ denotes the weighted brightness for the hot
gas or member galaxies at the $i$th pixel, here $(\vec{x}_i,
\vec{y}_i)$ is the relative coordinate of the $i$th pixel to the
cluster center. For the 24 quasi-equilibrium clusters in our sample,
we separately take the X-ray brightness peak as the center for the ICM
and the position of BCG as the center for member galaxies, though the
deviations between the X-ray peak and the BCG are small for these
quasi-equilibrium clusters, see Fig.~\ref{fig1}. Briefly
speaking, the $\vec{\eta}_{\rm gal}$ for member galaxies indicates the
normalized offset between the BCG and the optical flux-weighted
centroid, while the $\vec{\eta}_{\rm ICM}$ means the departure from
the X-ray brightness peak to the X-ray flux-weighted centroid.

\begin{table*}
\centering
\rotatebox[origin=c]{90}{%
\begin{varwidth}{\textheight}
\caption{Parameters for the 24 clusters in  quasi-equilibrium state.}
\tabcolsep=1.8pt
\footnotesize
\begin{tabular}{lrrcccccrrccrrrccr}
\hline
  \multicolumn{1}{c}{Name} &\multicolumn{1}{c}{RA}  &\multicolumn{1}{c}{DEC} & $z$   &  \multicolumn{1}{c}{log$_{10}(c)$} & $r_{500}$ & \multicolumn{1}{c}{$M_{500}$} &\multicolumn{1}{c}{$N_{\rm sat}$}  &  \multicolumn{2}{c}{$\vec{\eta}_{\rm ICM}$}  &  \multicolumn{1}{c}{$\gamma_{\rm ICM}$}  &  \multicolumn{1}{c}{$e_{\rm ICM}$} &  \multicolumn{1}{c}{$\phi_{\rm ICM}$}  &  \multicolumn{2}{c}{$\vec{\eta}_{\rm gal}$} & \multicolumn{1}{c}{$\gamma_{\rm gal}$} & \multicolumn{1}{c}{$e_{\rm gal}$} & \multicolumn{1}{c}{$\phi_{\rm gal}$}\\
\cline{9-10} \cline{14-15}
 & \multicolumn{1}{c}{(J2000)}& \multicolumn{1}{c}{(J2000)} & & & \multicolumn{1}{c}{(Mpc)} &\multicolumn{1}{c}{($10^{14}M_{\odot}$)} & &\multicolumn{1}{c}{(X-axis, $\%$)} &\multicolumn{1}{c}{(Y-axis, $\%$)} &\multicolumn{1}{c}{($\%$)} & &\multicolumn{1}{c}{($^{\circ}$)} &\multicolumn{1}{c}{(X-axis, $\%$)} &\multicolumn{1}{c}{(Y-axis, $\%$)} &\multicolumn{1}{c}{($\%$)} & &\multicolumn{1}{c}{($^{\circ}$)} \\
  \multicolumn{1}{c}{(1)} &\multicolumn{1}{c}{(2)}  &\multicolumn{1}{c}{(3)} & \multicolumn{1}{c}{(4)}   &  \multicolumn{1}{c}{(5)} & \multicolumn{1}{c}{(6)} & \multicolumn{1}{c}{(7)} & \multicolumn{1}{c}{(8)} & \multicolumn{1}{c}{(9)} & \multicolumn{1}{c}{(10)} & \multicolumn{1}{c}{(11)} & \multicolumn{1}{c}{(12)} & \multicolumn{1}{c}{(13)} & \multicolumn{1}{c}{(14)} & \multicolumn{1}{c}{(15)} & \multicolumn{1}{c}{(16)} & \multicolumn{1}{c}{(17)} & \multicolumn{1}{c}{(18)} \\
  \hline                    
A85                    & 10.46029 &-9.30313 &0.0552 &-0.42$\pm$0.01 &1.21 & 5.32 &31 & 0.72$\pm$0.20 &-1.36$\pm$0.23 &29.3$\pm$8.0 &0.07$\pm$0.01 &170.7$\pm$0.3 & 10.26$\pm$4.24 & -5.38$\pm$6.97 &48.3$\pm$7.7  &0.21$\pm$0.01 &174.3$\pm$5.9 \\
A291                   & 30.42959 &-2.19669 &0.1963 &-0.31$\pm$0.01 &1.08 & 4.31 &26 &-0.33$\pm$0.20 & 0.09$\pm$0.20 &26.3$\pm$7.4 &0.03$\pm$0.01 &140.3$\pm$0.3 & -1.17$\pm$4.91 &  7.38$\pm$3.00 &45.2$\pm$5.9  &0.37$\pm$0.01 &112.8$\pm$0.4 \\
A383                   & 42.01413 &-3.52923 &0.1889 &-0.29$\pm$0.01 &1.07 & 4.25 &35 & 1.10$\pm$0.19 &-0.16$\pm$0.20 &26.5$\pm$8.4 &0.01$\pm$0.01 & 13.4$\pm$1.1 &  0.65$\pm$4.62 & -2.22$\pm$5.08 &43.2$\pm$9.0  &0.34$\pm$0.06 & 33.2$\pm$0.6 \\
A586                   &113.08453 &31.63353 &0.1875 &-0.49$\pm$0.02 &1.16 & 5.20 &32 & 0.74$\pm$0.21 & 0.67$\pm$0.21 &24.5$\pm$6.6 &0.02$\pm$0.01 &131.7$\pm$0.3 &  0.28$\pm$4.41 &  2.61$\pm$4.43 &33.1$\pm$4.5  &0.12$\pm$0.01 &153.2$\pm$2.3 \\
A598                   &117.85459 &17.51422 &0.1865 &-0.30$\pm$0.01 &1.04 & 3.82 &26 & 1.12$\pm$0.20 & 1.91$\pm$0.19 &25.4$\pm$5.4 &0.03$\pm$0.01 &103.6$\pm$0.1 &  8.40$\pm$7.57 &  6.63$\pm$3.70 &36.6$\pm$5.6  &0.39$\pm$0.01 & 99.7$\pm$0.3 \\
A795                   &141.02211 &14.17263 &0.1356 &-0.42$\pm$0.01 &1.05 & 3.81 &40 & 0.59$\pm$0.23 &-1.12$\pm$0.21 &27.2$\pm$8.2 &0.07$\pm$0.01 &113.2$\pm$0.1 & -9.89$\pm$4.86 &-13.24$\pm$5.59 &60.7$\pm$7.0  &0.30$\pm$0.01 &141.6$\pm$0.8 \\
ZwCl1021.0+0426        &155.91515 & 4.18629 &0.2897 &-0.29$\pm$0.01 &1.37 & 9.17 &31 & 1.79$\pm$0.18 & 1.40$\pm$0.19 &24.4$\pm$4.8 &0.03$\pm$0.01 &146.0$\pm$0.3 &  3.79$\pm$3.07 & -0.39$\pm$6.65 &44.1$\pm$5.8  &0.41$\pm$0.05 & 19.0$\pm$0.6 \\
ZwCl1023.3+1257        &156.49161 &12.68566 &0.1420 &-0.37$\pm$0.01 &1.02 & 3.48 &21 & 0.06$\pm$0.20 & 1.03$\pm$0.22 &25.3$\pm$5.6 &0.06$\pm$0.01 &158.0$\pm$0.2 &  3.21$\pm$4.11 &  1.08$\pm$4.58 &42.2$\pm$8.2  &0.23$\pm$0.03 & 12.9$\pm$1.7 \\
A1650                  &194.67288 &-1.76146 &0.1200 &-0.44$\pm$0.01 &1.10 & 4.12 &25 & 0.81$\pm$0.21 & 1.40$\pm$0.23 &24.1$\pm$5.1 &0.06$\pm$0.01 &159.8$\pm$0.2 & -0.28$\pm$4.01 &  3.63$\pm$6.09 &47.3$\pm$9.2  &0.33$\pm$0.02 &  5.4$\pm$1.0 \\
A1689                  &197.87296 &-1.34112 &0.1920 &-0.43$\pm$0.01 &1.35 & 8.39 &85 &-0.40$\pm$0.19 & 1.42$\pm$0.20 &28.1$\pm$6.6 &0.04$\pm$0.01 & 14.6$\pm$0.2 & -2.43$\pm$2.13 & -0.21$\pm$2.52 &34.6$\pm$6.4  &0.21$\pm$0.02 & 15.7$\pm$0.4 \\
A1795                  &207.21877 &26.59293 &0.0624 &-0.32$\pm$0.01 &1.22 & 5.53 &25 & 0.53$\pm$0.19 & 1.20$\pm$0.21 &26.8$\pm$6.8 &0.06$\pm$0.01 & 15.2$\pm$0.1 & -5.91$\pm$4.03 & -5.66$\pm$5.65 &49.6$\pm$6.8  &0.30$\pm$0.05 & 26.9$\pm$0.4 \\
A1835                  &210.25862 & 2.87847 &0.2520 &-0.30$\pm$0.01 &1.42 &10.50 &66 & 0.43$\pm$0.18 & 0.02$\pm$0.19 &25.9$\pm$7.6 &0.03$\pm$0.01 &168.8$\pm$0.5 & -1.54$\pm$2.59 &  1.83$\pm$3.40 &33.5$\pm$4.7  &0.21$\pm$0.01 &164.0$\pm$1.5 \\
A1902                  &215.41701 &37.29130 &0.1584 &-0.39$\pm$0.01 &1.04 & 3.80 &25 & 2.54$\pm$0.21 & 0.15$\pm$0.20 &28.9$\pm$8.7 &0.05$\pm$0.01 & 67.7$\pm$0.1 & -2.62$\pm$4.38 &-10.12$\pm$7.15 &50.7$\pm$5.7  &0.07$\pm$0.02 & 49.8$\pm$1.1 \\
ZwCl1454.8+2233        &224.31294 &22.34289 &0.2576 &-0.26$\pm$0.01 &1.20 & 6.32 &29 & 0.02$\pm$0.18 &-0.39$\pm$0.19 &27.0$\pm$8.6 &0.04$\pm$0.01 & 27.0$\pm$0.3 & -2.52$\pm$3.13 &  4.53$\pm$5.01 &35.3$\pm$9.0  &0.15$\pm$0.02 & 59.6$\pm$0.8 \\
A2009                  &225.08131 &21.36941 &0.1510 &-0.38$\pm$0.01 &1.15 & 5.06 &34 & 0.58$\pm$0.20 & 0.14$\pm$0.22 &26.3$\pm$7.2 &0.05$\pm$0.01 &163.1$\pm$0.3 & -0.89$\pm$5.30 & -1.93$\pm$2.39 &42.3$\pm$6.2  &0.37$\pm$0.01 &158.8$\pm$0.9 \\
RBS1460                &226.03130 &-2.80461 &0.2169 &-0.21$\pm$0.01 &1.52 &12.48 &41 &-0.52$\pm$0.18 &-0.33$\pm$0.17 &27.6$\pm$8.0 &0.05$\pm$0.01 & 58.8$\pm$0.1 & -0.90$\pm$3.35 & -3.44$\pm$4.09 &45.6$\pm$7.3  &0.47$\pm$0.02 & 85.4$\pm$0.1 \\
A2029                  &227.73376 & 5.74478 &0.0785 &-0.35$\pm$0.01 &1.33 & 7.27 &51 & 0.68$\pm$0.19 &-0.63$\pm$0.21 &28.1$\pm$8.9 &0.05$\pm$0.01 & 11.7$\pm$0.2 & -0.07$\pm$4.95 &  0.28$\pm$6.47 &45.1$\pm$9.6  &0.21$\pm$0.02 & 22.5$\pm$0.8 \\
A2052                  &229.18536 & 7.02162 &0.0345 &-0.28$\pm$0.01 &0.95 & 2.49 &23 &-0.10$\pm$0.21 &-0.68$\pm$0.22 &27.6$\pm$8.6 &0.05$\pm$0.01 & 33.1$\pm$0.2 &  3.85$\pm$2.87 &  3.48$\pm$4.48 &36.0$\pm$6.0  &0.27$\pm$0.01 &156.1$\pm$1.7 \\
A2063                  &230.77208 & 8.60922 &0.0341 &-0.46$\pm$0.01 &0.90 & 2.16 &33 &-0.16$\pm$0.23 & 0.97$\pm$0.23 &27.0$\pm$7.5 &0.02$\pm$0.01 & 44.2$\pm$0.4 &  0.76$\pm$3.33 & -0.21$\pm$4.07 &37.6$\pm$4.2  &0.23$\pm$0.01 &172.5$\pm$2.7 \\
A2199                  &247.15935 &39.55126 &0.0312 &-0.31$\pm$0.01 &1.00 & 2.96 &28 &-0.41$\pm$0.20 & 0.72$\pm$0.21 &27.7$\pm$7.6 &0.04$\pm$0.01 & 34.5$\pm$0.3 &  5.70$\pm$4.31 & -0.99$\pm$4.16 &40.4$\pm$6.4  &0.31$\pm$0.01 &152.0$\pm$0.7 \\
A2244                  &255.67705 &34.05999 &0.0993 &-0.45$\pm$0.01 &1.13 & 4.49 &38 & 0.23$\pm$0.20 &-0.37$\pm$0.21 &26.4$\pm$8.1 &0.02$\pm$0.01 &170.5$\pm$0.6 &  8.35$\pm$4.26 & -8.69$\pm$4.01 &40.1$\pm$8.8  &0.26$\pm$0.04 & 49.5$\pm$0.5 \\
RBS1639                &260.04184 &26.62557 &0.1604 &-0.30$\pm$0.01 &1.22 & 6.01 &31 &-0.09$\pm$0.19 &-0.04$\pm$0.21 &26.4$\pm$8.1 &0.05$\pm$0.01 & 17.8$\pm$0.2 & -1.21$\pm$5.12 &  4.01$\pm$6.63 &37.8$\pm$7.5  &0.15$\pm$0.03 & 40.9$\pm$0.7 \\
A2261                  &260.61325 &32.13257 &0.2233 &-0.43$\pm$0.01 &1.30 & 7.88 &48 & 1.43$\pm$0.21 &-0.27$\pm$0.20 &27.7$\pm$9.2 &0.04$\pm$0.01 & 75.5$\pm$0.1 & 17.16$\pm$4.46 &  1.08$\pm$3.74 &51.7$\pm$7.0  &0.26$\pm$0.02 & 98.9$\pm$0.2 \\
RBS1748                &322.41647 & 0.08921 &0.2339 &-0.36$\pm$0.01 &1.25 & 7.04 &30 & 1.64$\pm$0.21 &-0.67$\pm$0.19 &29.4$\pm$9.7 &0.06$\pm$0.01 & 69.0$\pm$0.1 &-12.57$\pm$5.40 & -0.29$\pm$3.72 &67.1$\pm$11.5 &0.49$\pm$0.05 & 68.1$\pm$0.2 \\
\hline                  
\end{tabular}
\label{tab1}      
    {Notes: Columns: (1) cluster name; (2 - 4) right ascension,
      declination and redshift of the BCG measured from the {\it SDSS}
      data; (5) the concentration index of the cluster, taken from
      \citet{yhw22}; (6 - 7) the radius and mass of the cluster, from
      \citet{pap+11}; (8) the number of member galaxies brighter than
      --20.0 mag in $0.5r_{500}$; (9 - 10) the normalized offset
      between the brightness peak and the estimated centroid from the
      cluster X-ray image, values in the X-axis and Y-axis are
      presented separately; (11 - 13) the sphere index, the
      ellipticity and the position angle calculated from the X-ray
      image; (14 - 15) the normalized offset between the BCG and the
      centroid estimated from member galxies in X-axis and Y-axis; (16 -
      18) the sphere index, the ellipticity and position angle
      estimated from the distribution of member galaxies.}
\end{varwidth}}
\end{table*}

The dynamical state of member galaxies and hot gas could be 
compared with the derived mass distribution, rather than their 
brightness distribution. In optical, the mass of galaxies is
approximately proportional to their luminosity \citep[e.g.,][]{fj76}. 
In X-ray, the brightness of hot gas is broadly proportional to the 
square of the in-situ mass \citep[e.g.,][]{pv08}. We calculate the 
dynamical parameters using the weighted brightness $f_{{\rm w},i}$, 
which is defined as
\begin{equation}
\begin{cases}
f_{{\rm w},i}=\sqrt{f_{i}}~~~~~~{\rm(for~ICM)},\\ 
f_{{\rm w},i}=f_{i}~~~~~~{\rm(for~galaxy)},\\ 
\end{cases}
\label{eqt00}
\end{equation}
where $f_{i}$ is the pixel value of X-ray images or the luminosity of
member galaxies.

Ideally, a relaxed galaxy cluster shows a spherically symmetric
morphology, thus the radial profiles of the matter distribution should be
similar in any directions. We define the sphere index $\gamma$ to
quantify the degree of spherical symmetry of the ICM and member
galaxies in clusters. First, we define a frame from the cluster center
with an arbitrary position angle $\theta$, and the two perpendicular
axes of the frame divide the image into four quadrants. Second, we
calculate the parameter $S_{\theta,n}$ as the sum of all pixels in the
$n$th quadrant as
\begin{equation}
\label{eqt01}
 S_{\theta,n}=\sum\limits_{i}f_{{\rm w},i}\cdot r_{i},
\end{equation}
the $r_i$ is the distance from the $i$th pixel to the cluster center,
and $n=1,2,3,4$. Then, the $\Delta S_{\theta}$ is defined as
\begin{equation}
\label{eqt02}
 \Delta S_{\theta}=S_{\theta,{\rm max}}-S_{\theta,{\rm min}},
\end{equation}
where $S_{\theta,{\rm max}}$ $(S_{\theta,{\rm min}})$ is the maximum
(minimum) among the
$(S_{\theta,1},~S_{\theta,2},~S_{\theta,3},~S_{\theta,4})$. Last, we
search for the $\theta$, that yields a maximum value for $\Delta S_{\theta}$, 
and the sphere index $\gamma$ is defined as
\begin{equation}
\label{eqt03}
 \gamma=\frac{{\rm max}\left(\Delta S_{\theta}\right)}{S_{\rm tot}}=\frac{{\rm max}\left(\Delta S_{\theta}\right)}{\sum\limits_{n=1}^{4} S_{\theta,n}}\times100\%.
\end{equation}
In short, the $\gamma$ indicates the largest difference in matter
distribution between any two quadrants. 
More relaxed clusters are expected to have a smaller $\gamma$.

The ellipticity and orientation of galaxy clusters has been calculated
with different methods \citep[e.g.,][]{cm80,nsd+10,hmf+16,yw22}. To
avoid a possible bias caused by different methods, here cluster
ellipticities and orientations from optical and X-ray data are
calculated in the same way. We choose the moments method, which has
been widely used for optical data, but adopt the new weight of
brightness $f_{{\rm w},i}$ rather than the traditional weight of
$\frac{1}{r_i^2}$ \citep[e.g.,][]{nsd+10,hmf+16} to make the method
also applicable to diffuse X-ray images. The second-order moments are
defined as
\begin{equation}
\begin{split}
& \mu_{20}=\frac{1}{\sum\limits_{i}f_{{\rm w},i}}\sum\limits_{i}f_{{\rm w},i}\cdot x_{i}^2 ,\\
& \mu_{02}=\frac{1}{\sum\limits_{i}f_{{\rm w},i}}\sum\limits_{i}f_{{\rm w},i}\cdot y_{i}^2,\\
& \mu_{11}=\frac{1}{\sum\limits_{i}f_{{\rm w},i}}\sum\limits_{i}f_{{\rm w},i}\cdot x_{i}y_{i}.
\end{split}
\end{equation}
The major and minor axes of the cluster, $\Lambda_{\rm a}$ and
$\Lambda_{\rm b}$, can be computed through the following equation
\citep[e.g.,][]{cm80}
\begin{equation}
\begin{vmatrix}
\label{eqt2}
 \mu_{20}-\Lambda^2 & \mu_{11}\\ 
 \mu_{11} & \mu_{02}-\Lambda^2\\
\end{vmatrix}=0.
\end{equation}
 The cluster ellipticity is defined as
\begin{equation}
\label{eqt3}
 e=1-\frac{\Lambda_{\rm b}}{\Lambda_{\rm a}}.
\end{equation}
The position angle of the cluster is calculated with
\begin{equation}
\label{eqt4}
\phi={\rm arctan}\left(\frac{\mu_{02}-\Lambda_{\rm a}^{2}}{\mu_{11}}\right),
\end{equation}
here the $\phi$ is from the north through the east. The deviation
between position angles measured from the distributions of member
galaxies ($\phi_{\rm gal}$) and ICM ($\phi_{\rm ICM}$) is denoted as
\begin{equation}
\begin{cases}
\Delta\phi=|\phi_{\rm gal}-\phi_{\rm ICM}|~~~~~~~~~~~~~~~~(|\phi_{\rm
  gal}-\phi_{\rm
  ICM}|\le90^{\circ}),\\ \Delta\phi=180^{\circ}-|\phi_{\rm
  gal}-\phi_{\rm ICM}|~~~(|\phi_{\rm gal}-\phi_{\rm ICM}|>90^{\circ}).
\end{cases}
\label{Phio}
\end{equation}
Here $\Delta\phi$ is in the range of
$0^{\circ}\le\Delta\phi\le90^{\circ}$.

Considering the possible ellipticity changes at different cluster
radii \citep[e.g.,][]{sml+97} and also biases for other parameters, we
calculate the normalized centroid offsets, sphere indices, ellipticities and position
angles from optical and X-ray images in the same
region encircled by 0.5$r_{500}$, see the large dotted circles in
Fig.~\ref{fig1}.

\begin{figure}
\centering
\includegraphics[width=0.35\textwidth, angle=-90]{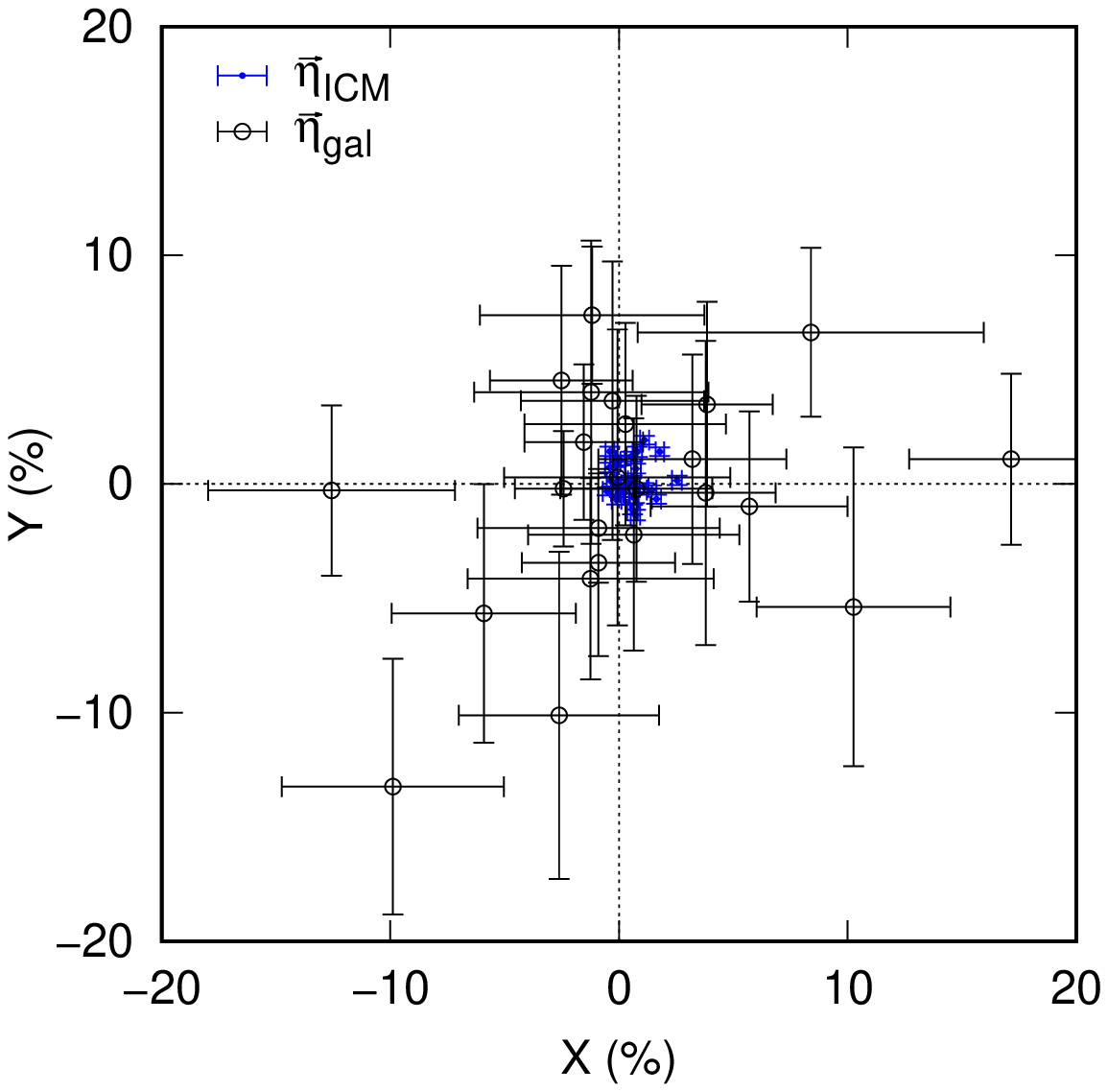}
\caption{The distribution of the normalized centroid offsets estimated
  from the ICM ($\vec{\eta}_{\rm ICM}$) and member galaxies
  ($\vec{\eta}_{\rm gal}$). The intersection of the dotted lines
  indicate the position of the X-ray brightness peak for
  $\vec{\eta}_{\rm ICM}$ and the position of BCG for $\vec{\eta}_{\rm
    gal}$. The errorbars of each point are the Poisson errors.}
\label{fig2}
\end{figure}

\section{Different relaxed state for member galaxies and ICM}

The exact dynamical state should be derived from the 3 dimensional 
velocities of all kinds of matter in a cluster. However, the velocities 
of dark matter and intracluster gas are not available. For galaxies, 
only the velocity along the line of sight can be well measured by the 
spectroscopic redshifts, which are available only for a small fraction 
of members. Therefore, in this paper, we only consider the mass 
distribution in the sky plane for the intracluster gas and member galaxies.
Clear displacements between the ICM and member galaxies have been
observed in many merging clusters
\citep[e.g.,][]{mcf+08,dwj+12,jhm+14,gdw+16,gvd+17}, as member galaxies, similar to the dark
matter, move almost collisionless during the cluster merger, while the
intracluster gas evolves under dynamic pressures \citep[e.g.,][]{pkb09}. The simulation made by \citet{pfb+06}
indicates that the dark matter generally takes longer time to reach
the virialization state than the gaseous component, especially for
mergers between subclusters with large mass difference (e.g.,
$m_1:m_2<1:3$). Considering that both member galaxies and the dark
matter are almost collisionless, it is natural to expect the
relaxation time for member galaxies is also longer than that of the
ICM. In the following, we show the different relaxed state of member
galaxies and ICM in the 24 quasi-equilibrium clusters.

The departure of the flux-weighted centroid from the brightness peak is
generally very small for symmetric distributions, but could be large
for disturbed systems. In Fig.~\ref{fig2}, the distribution of
normalized centroid offsets is shown for the ICM
($\vec{\eta}_{\rm ICM}$, dots) and member galaxies ($\vec{\eta}_{\rm
  gal}$, circles). It is clear that the offsets calculated from
  member galaxies ($\langle\vec{\eta}_{\rm gal}\rangle=6.11\%~
  r_{500}$) are more scattered than those from the ICM
  ($\langle\vec{\eta}_{\rm ICM}\rangle=1.12\%~r_{500}$).

\begin{figure}
\centering
\includegraphics[width=0.35\textwidth, angle=-90]{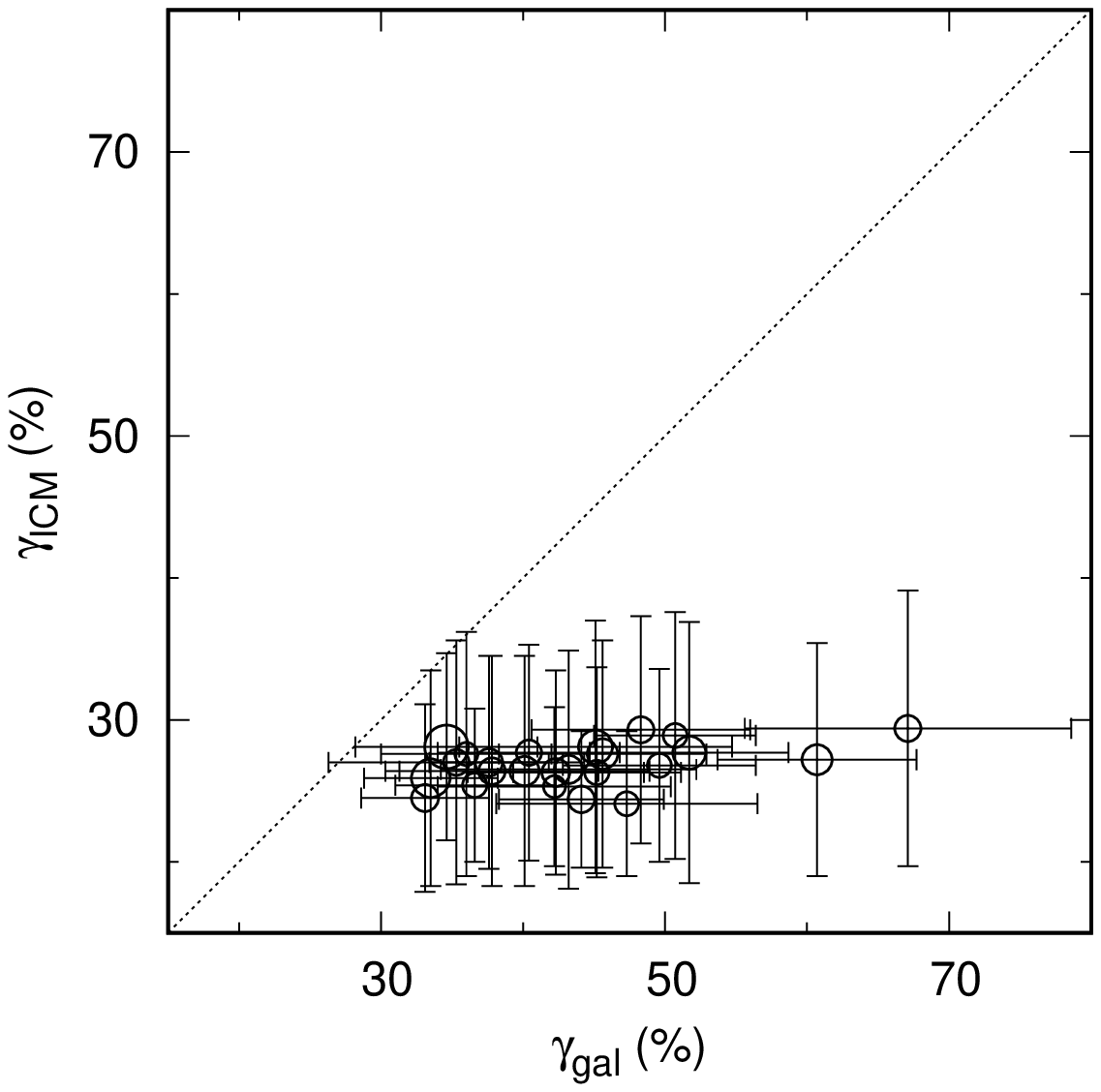}\\[2mm]
\includegraphics[width=0.35\textwidth, angle=-90]{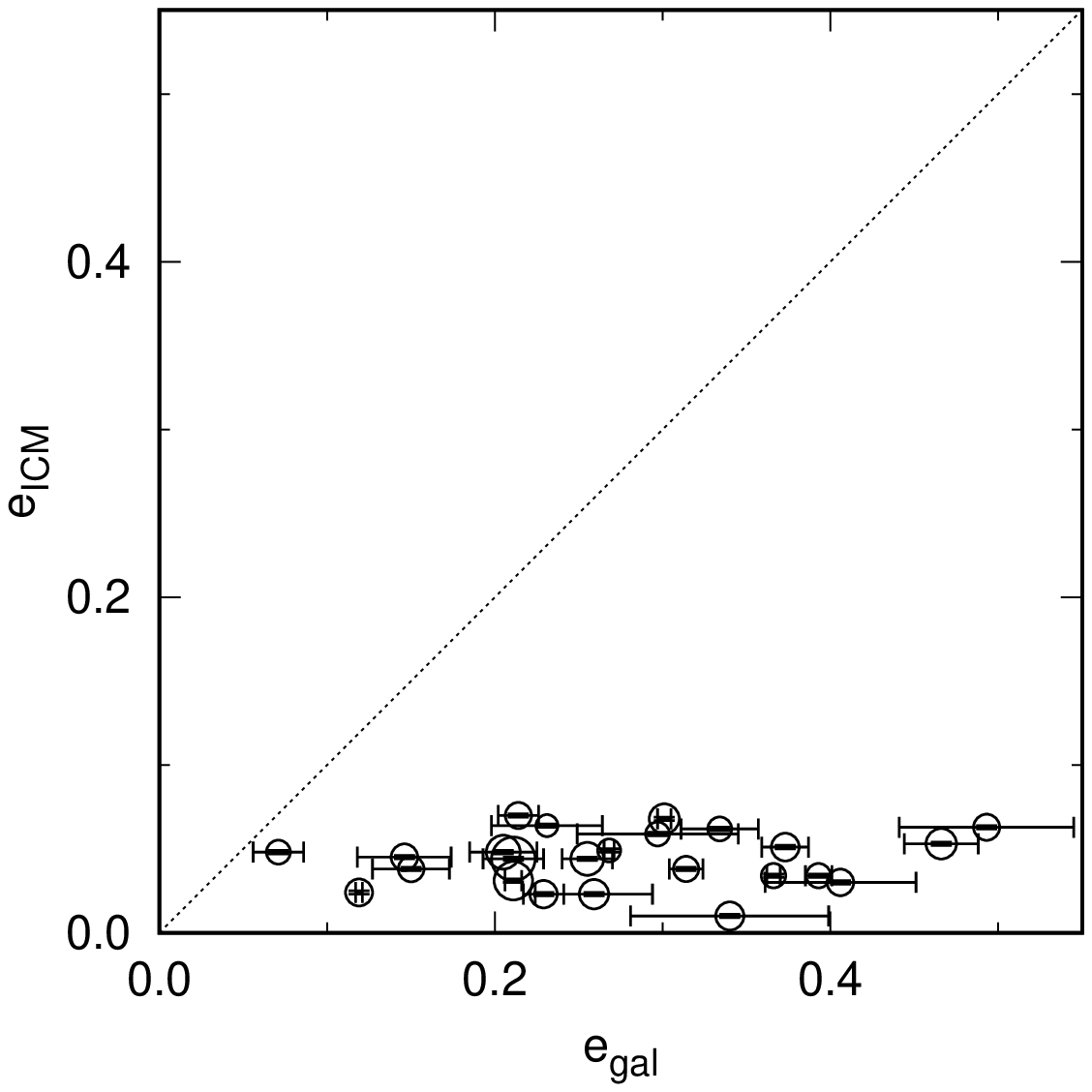}
\caption{Comparisons of the sphere indices (upper panel) and
  ellipticities (lower panel) of clusters measured from member
  galaxies and ICM. The circle size is proportional to $\sqrt{N_{\rm
      sat}}$. The dotted lines in both panels show the equivalence for
  the X-axis and Y-axis. The errorbars in the two panels are the
  Poisson errors.}
\label{fig3}
\end{figure}

Observationally, galaxy clusters in dynamical equilibrium state
usually show a roundish morphology with a small difference in any
direction. In Fig.~\ref{fig3}, we compare the sphere indices (upper
panel) and ellipticities (lower panel) of clusters measured from
distributions of member galaxies and ICM. The upper panel
shows that all the 24 clusters exhibit larger
non-spherical symmetry for member galaxies than for the ICM. The lower
panel indicates that all clusters show larger $e_{\rm gal}$,
suggesting that the distribution of member galaxies is more elongated
than that of the diffuse ICM. If we take galaxies as
the tracer of dark matter, our result is consistent with the
simulations made by \citet{mog22}, which show that $\sim$90\% clusters
display a more elongated dark matter morphology when compared to the
gaseous component.

\begin{figure}
\centering
\includegraphics[width=0.35\textwidth, angle=0]{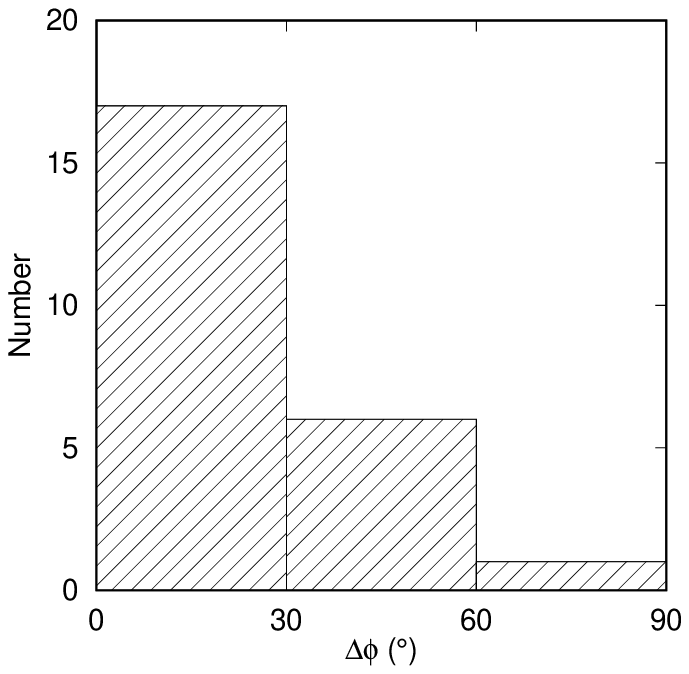}
\caption{The distribution of $\Delta\phi$ for the 24 clusters,
  $\Delta\phi$ is the difference between cluster orientations
  estimated from member galaxies and the ICM.}
\label{fig4}
\end{figure}

In Fig.~\ref{fig4}, we show the distribution of $\Delta\phi$, the
deviation angle between cluster orientations estimated from member
galaxies and the ICM, for the 24 clusters. It is clear that the
cluster orientations indicated by member galaxies tend to align with
those measured from the ICM, which is consistent with our recent work
\citep{yw22}. The orientation uncertainties are generally very small,
see Table~\ref{tab1}, so the alignment will not be significantly 
weakened by data uncertainties. The spatial orientation is
the reflection of cluster angular momentum, which is further linked to
the direction(s) of the post merger(s). The orientation alignment
shown in Fig.~\ref{fig4} indicates that the information of post
mergers is stored simultaneously in the distributions of member
galaxies and ICM, though the structures of the two components 
differ in detail.

The large discrepancy between the ICM and member galaxies in clusters,
like in the Bullet cluster, indicates that they evolve differently in the
early stages of merging. The results presented in this paper describe the 
picture of dynamic evolution in later stages. The angular momentum accompanied by the
post-merger is stored in or transformed into the ICM and member
galaxies simultaneously, therefore the cluster orientations indicated
by the two matter components are consistent with each other. The
diffuse ICM evolves under non-negligible viscosity, causing the angular
momentum of the ICM introduced by the merger to dissipate quickly
into other directions. As a result, the distribution of the ICM tends to 
be isotropic, giving the ICM a roundish morphology with small
centroid offset and asymmetry. However, collisions between member
galaxies are rare, so the relaxation timescale of member
galaxies is larger than that of the ICM. Such cases have been simulated 
by \citet{pfb+06}. For two clusters with a mass ratio of
1:1, the relaxation timescale for the whole system is about 
$t_{\rm relax}\sim 4.4-5.4$~Gyr. The gaseous component generally 
reaches the virialization state about $\sim$0.5~Gyr eariler than 
the dark matter. For the 10:1 mergers, the virialization timescale of the 
gaseous components is about 6.4 -- 7.4 Gyr, while the dark matter 
needs much longer time ($t>14$ Gyr) to reach the virialization 
criteria in $r_{200}$.

\begin{figure*}
\centering
\includegraphics[width=0.65\textwidth, angle=0]{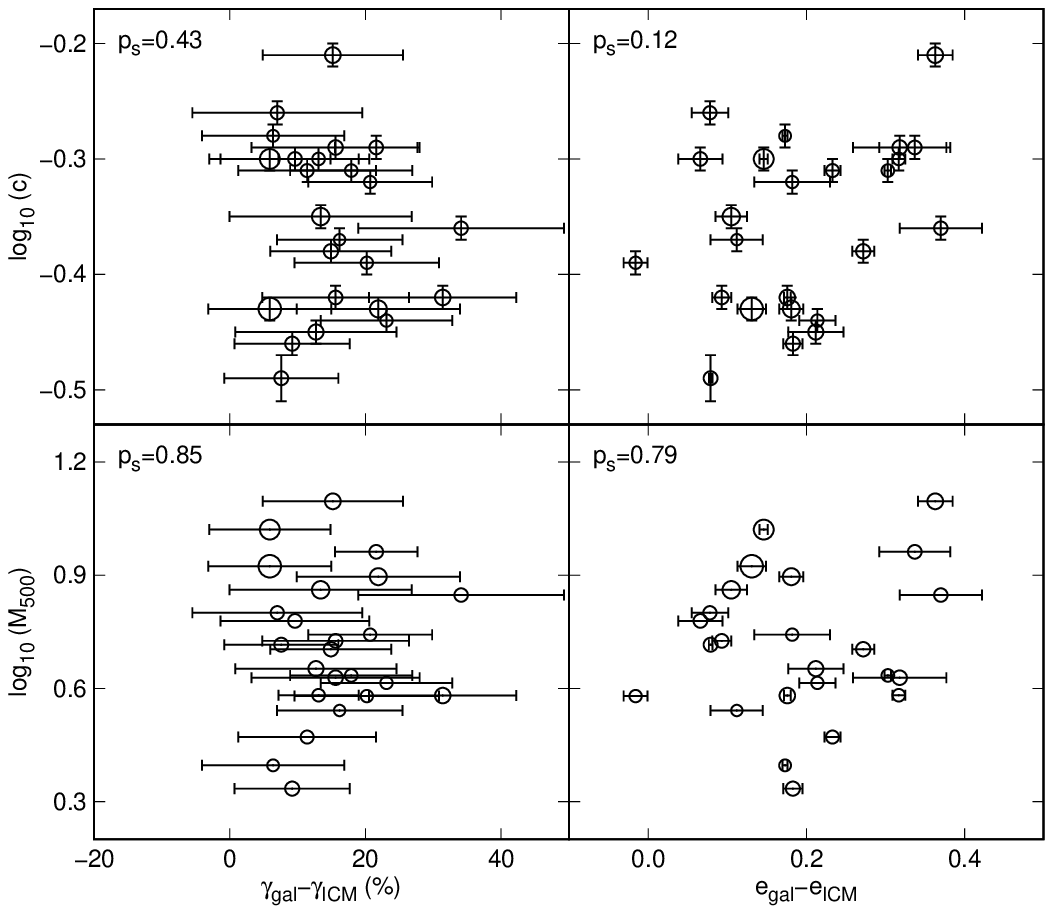}
\caption{Left-hand panels: relations between the $\gamma_{\rm
    gal}-\gamma_{\rm ICM}$, the difference of sphere index measured
  from distributions of member galaxies and the ICM, against the
  concentration index $c$ (upper panel) and the cluster mass $M_{500}$
  (lower panel). The circle size is proportional to $\sqrt{N_{\rm
      sat}}$. The errorbars of each point are the Poisson errors. The
  significance of Spearman rank-order correlation $p_{\rm s}$ is
  written on the top-left corner of each panel. Right-hand panels:
  similar to the left-hand panels but for the ellipticity of galaxy
  clusters.}
\label{fig5}
\end{figure*}

In Fig.~\ref{fig5}, we explore if the difference of dynamic features
between member galaxies and the ICM is related to other properties of
clusters, i.e., the concentration index $c$ (upper panels) and the
cluster mass $M_{500}$ (lower panels). The significance of the
Spearman rank-order correlation $p_{\rm s}$ \citep[see definition in
  p. 640 of][]{ptv+92} is used to assess weak but
intrinsic correlations, with a zero value for a robust correlation
and a non-zero value for no correlations. All four panels show a
clear non-zero $p_{\rm s}$, which indicates that the difference in
dynamical features for member galaxies and the ICM is independent of
cluster mass and concentration index.

\section{Summary}
Earlier studies have shown that member galaxies and the ICM in galaxy 
clusters may exhibit different dynamically relaxed states, especially in the early stages of a cluster merger
\citep[e.g.,][]{bat+08}. Simulations suggest that the almost
collisionless dark matter and member galaxies generally shows more
elongated distribution \citep{mog22} and take longer time to reach the
dynamical equilibrium state \citep[e.g.,][]{pfb+06} than the diffuse
gaseous component. In this study, we focus on 24 massive clusters that are in quasi-equilibrium
state as indicated by their X-ray images. The cluster orientations and
three kinds of dynamical parameters, i.e., the normalized centroid
offset $\eta$, the sphere index $\gamma$ and the ellipticity $e$, for
the distributions of member galaxies and ICM are calculated through
the same method in the same region. While the cluster orientations
indicated by the two matter components are well aligned, the
dynamical parameters derived from member galaxies are systematically
larger than those obtained from the diffuse ICM. This suggests that the
diffuse gaseous component reaches a more relaxed state than 
collisionless member galaxies after cluster mergers. We find that the difference in
dynamical features between member galaxies and the ICM is not related
to cluster mass and center concentration.

\section*{Acknowledgements}
We thank the anonymous referee for insightful comments which
  helped us to improve the paper.
This work is partially supported by the National Natural Science Foundation
of China (No. 11988101, 12073036, 11833009), the science research grants from the China Manned Space Project (No. CMS-CSST-2021-A01,
CMS-CSST-2021-B01) and the National SKA Program of China (Grant
No. 2022SKA0120103).
HB obtains support from the Deutsche Forschungsgemeinschaft through the Excellence Cluster Origins and GC is supported by the Deutsches Zentrum f\"ur Luft- und Raumfahrt through project 50 OR 2204.
This work is based on observations obtained with {\it XMM-Newton}, an
ESA science mission with instruments and contributions directly funded
by ESA Member States and NASA. 
Funding for the Sloan Digital Sky Survey IV has been provided by the
Alfred P. Sloan Foundation, the U.S. Department of Energy Office of
Science, and the Participating Institutions. SDSS acknowledges support
and resources from the Center for High-Performance Computing at the
University of Utah. The SDSS web site is www.sdss.org.

\section*{Data availability}

The data underlying this article will be shared on reasonable request 
to the first author. More optical and X-ray data on large samples of 
galaxy clusters can be found at the webpage: http://zmtt.bao.ac.cn/galaxy\_clusters/.

\bibliographystyle{mnras}
\bibliography{ref}

\label{lastpage}
\end{document}